\documentclass[prd,preprintnumbers,nofootinbib,11pt]{article}

\usepackage{verbatim} 
\usepackage{graphicx,amsmath,amssymb,epsfig}
\def\beq{\begin{equation}}
\def\eeq{\end{equation}}

\def\beqn{\begin{eqnarray}}
\def\eeqn{\end{eqnarray}}

\begin{document}
\title{\Large \bf 
 Integrable spin chains and scattering amplitudes}
\author{\large J. Bartels$^1$, L.~N. Lipatov$^{1,2}$
and  A. Prygarin$^1$
\\
{\it  
$^1$~
II. Institut f\"{u}r Theoretische Physik, Universit\"{a}t Hamburg, Germany}\\
{\it
$^2$~
Petersburg Nuclear Physics Institute and Petersburg State University, Russia}}

\maketitle

\vspace{-7cm}
\begin{flushright}
{\small DESY-11-051}
\end{flushright}
\vspace{7cm}
\begin{abstract}
\noindent
In this review we show that the multi-particle scattering amplitudes in $\mathcal{N}=4$
SYM at large $N_c$ and in the
multi-Regge kinematics for some physical regions have the high
energy behavior appearing from the contribution of the Mandelstam cuts
in the complex angular momentum plane of  the corresponding $t$-channel partial waves.  These Mandelstam cuts or Regge cuts are resulting from  gluon composite states  in the adjoint
representation of the gauge group $SU(N_c)$. In the leading logarithmic 
approximation (LLA) their contribution to the six point amplitude is in full 
agreement with the known two-loop result.  
 The Hamiltonian for
the Mandelstam states constructed from $n$ gluons in LLA coincides with the
local Hamiltonian of an integrable open spin chain. We construct the
corresponding wave functions using the integrals of motion and the
Baxter-Sklyanin approach.

This article is an invited review for a special issue of Journal of
Physics A devoted
 to Scattering Amplitudes in Gauge Theories.

\end{abstract}

\section{Introduction}

At high energies $s\gg -t$ in QCD the elastic scattering amplitude
for the process $AB\rightarrow A'B'$ in the
leading logarithmic approximation (LLA)
\beq
\alpha _s \,\ln s \sim 1\,,\,\,\alpha _s \ll 1
\eeq
has the Regge form~\cite{BFKL}
\begin{equation}
A_{2\rightarrow 2}=2\,g\delta _{\lambda _{A}\lambda _{A'}}
T_{AA'}^c\frac{s^{1+\omega (t)}}{t}\,g\,T_{BB'}^c
\,\delta _{\lambda _{B}\lambda _{B'}}\,,\,\,t=-\mathbf{q}^{{2}}\,.
\end{equation}
Here $T^c$ are the generators of the gauge group $SU(N_c)$,
$\lambda _r$ are the particle helicities and
$j(t)=1+\omega (t)$ is the gluon Regge trajectory
for the space-time dimension $D=4-2\epsilon$
\begin{equation}\label{reggetraj}
\omega (-\mathbf{q}^{2})=-\frac{\alpha_{s} N_c}{(2\pi )^2}\,
(2\pi \mu )^{2\epsilon}\,\int
d^{2-2\epsilon }\mathbf{k}
\,\frac{
\mathbf{q}^{2}}{\mathbf{k}^{2}(\mathbf{q}-\mathbf{k})^{2}}\approx
-\,a\,\left(\ln
\frac{\mathbf{q}^{2}}{\mu ^2}-\frac{1}{\epsilon}\right)\,.
\end{equation}
In the framework of the dimensional regularization
the parameter $\mu$ is the
renormalization point for the 't Hooft coupling constant
\begin{equation}
a=\frac{\alpha _{s}\,N_c}{2\pi }\,,
\end{equation}
where $\gamma =-\psi (1)$ is the Euler constant and
$\psi (x)=(\ln \Gamma (x))'$.
The gluon trajectory $j(t)$ was calculated also in the next-to-leading
approximation in QCD~\cite{trajQCD} and in the SUSY gauge
models~\cite{trajN4}.

In LLA gluons with momenta $k_r$ (r=1,..,n) are produced in the
multi-Regge kinematics
\begin{equation}
s=(p_A+p_B)^2\gg s_r=(k_r+k_{r-1})^2\gg -t_r=
{\bf q}_r^2\,,\,\,k_r=q_{r+1}-q_r\,,
\label{multReg}
\end{equation}
where the amplitude has the factorized form (see also section~\ref{JB})
\begin{eqnarray}
A_{2\rightarrow 2+n} =
2\,s\,\delta _{\lambda _{A}\lambda _{A'}}
g \, T^{c_1}_{AA'}
\frac{s_1^{\omega (-\vec{q}_1^2)}}{\vec{q}_1^2}gC_{\mu}(q_2,q_1)
e^*_\mu (k_1)T^{d_1}_{c_2c_1}\frac{s_2^{\omega
(-\vec{q}_2^2)}}{\vec{q}_2^2}
...\frac{s_{n+1}^{\omega (-\vec{q}_{n+1}^2)}}{\vec{q}_{n+1}^2}
\,g\,T^{c_{n+1}}_{BB'}\,\delta _{\lambda _{}\lambda _{B'}}\,.
\end{eqnarray} 
Here $C_\mu (q_2,q_1)$ is the
effective Reggeon-Reggeon-gluon vertex. In the case when the polarization
vector $e_{\mu}(k_1)$ describes the gluon with a positive
helicity in its c.m. system with the particle $A'$
one can obtain~\cite{effmult}
\begin{equation}
C\equiv C_\mu
(q_2,q_1)\,e^*_{\mu}(k_1)=\sqrt{2}\,\frac{q_2^*q_1}{k_1}\,,
\label{helicityproduction}
\end{equation}
where the complex notation $q=q_x+iq_y$ for the two-dimensional transverse
vector $\mathbf{q}$ was used.

The elastic scattering amplitude with vacuum quantum numbers in the
$t$-channel is calculated in terms of
the production amplitude $A_{2\rightarrow 2+n}$ with the
use of the $s$-channel
unitarity~\cite{BFKL}. In this approach the Pomeron appears as a composite
state of two Reggeized gluons. It is convenient to present the gluon
transverse coordinates in the complex form together with their canonically
conjugated momenta~\cite{effmult, int1}
\begin{equation}
\rho
_{k}=x_{k}+iy_{k}\,,\,\,\rho
_{k}^{\ast
}=x_{k}-iy_{k}\,,\,\,p_{k}=i
\frac{\partial }{\partial \rho
_{k}}\,,\,\,p_{k}^{\ast }=
i\frac{\partial }{\partial \rho
_{k}^{\ast }}\,.
\end{equation}
In this case the homogeneous Balitsky-Fadin-Kuraev-Lipatov (BFKL)
equation for the Pomeron wave function can be written as
follows~\cite{BFKL}
\begin{equation}
E\,\Psi
(\vec{\rho}_{1},\vec{\rho}_{2})=
H_{12}\,\Psi (\vec{\rho}_{1},\vec{%
\rho}_{2})\;,\,\,\Delta
=-\frac{\alpha _{s}N_{c}}{2\pi
}\,\min \,E\,,
\end{equation}
where $\Delta$ is the Pomeron intercept entering in the asymptotic
expression
for the total cross-section $\sigma _t\sim s^{\Delta}$.
The BFKL Hamiltonian has a
rather simple operator representation~\cite{int1}
\begin{equation}
H_{12}=\ln
\,|p_{1}p_{2}|^{2}+\frac{1}{p_{1}p_{2}^{\ast
}}(\ln \,|\rho _{12}|^{2})\,p_{1}p_{2}^{\ast
}+\frac{1}{p_{1}^{\ast
}p_{2}}(\ln \,|\rho
_{12}|^{2})\,p_{1}^{\ast
}p_{2}-4\psi (1)
\label{H12}
\end{equation}
with $\rho _{12}=\rho _1-\rho
_2$. The kinetic energy is proportional to the sum of two gluon Regge
trajectories $\omega (-|p_i|^2)$ ($i=1, 2$).
The potential energy $\sim \ln \,|\rho_{12}|^{2}$ is obtained by the
Fourier transformation from the product of
two gluon production vertices $C_\mu$. This Hamiltonian is invariant under
the
M\"{o}bius transformation~\cite{moeb}
\begin{equation}
\rho _{k}\rightarrow
\frac{a\rho _{k}+b}{c\rho
_{k}+d}\,,
\end{equation}
where $a,b,c$ and $d$ are complex parameters. The eigenvalues of the
corresponding Casimir operators are expressed in terms of the conformal
weights
\begin{equation}
m=\frac{1}{2}+i\nu
+\frac{n}{2}\,,\,\,\widetilde{m}=\frac{1}{2}+i\nu
-\frac{n}{2}\,,
\end{equation}
where $\nu$ and $n$ are respectively real and integer numbers
for the principal series of unitary
representations of the M\"{o}bius group $SL(2,C)$.
The eigenvalues of $H_{12}$ depend
on these parameters and can be written in the holomorphically
separable form~\cite{separ}
\beq
E_{m,\widetilde{m}}=\epsilon _m+\epsilon _{\widetilde{m}}\,,\,\,
\epsilon_m=\psi (m)+\psi (1-m)-2\psi (1)\,,
\label{PomE}
\eeq
where $\psi (x)=(\ln  \Gamma (x))'$\,.

The Pomeron intercept in LLA is positive
\beq
\Delta =4\;\frac{\alpha
_{s}}{\pi }N_{c}\,\ln 2>0
\eeq
and therefore the Froissart
bound $\sigma _t<c\ln ^2s$
for the total cross-section
is violated~\cite{BFKL}.
To restore the broken $s$-channel unitarity one
should take into account the contributions of diagrams corresponding to
the $t$-channel exchange
of an arbitrary number of reggeized gluons in the $t$-channel.
The wave function of the colorless state constructed from $n$ reggeized
gluons can be obtained in LLA as a solution of the
Bartels-Kwiecinski-Praszalowicz (BKP) equation~\cite{BKP}
\beq
E\,\Psi =H^{(0)}\,\Psi \,,\,\,\Delta
=-\frac{\alpha _{s}N_{c}}{4\pi}\,\min \,E.
\eeq
In the $N_c\rightarrow \infty$ limit the color structure is
simplified and the corresponding Hamiltonian has
the property of the
holomorphic separability~\cite{separ}
\begin{equation}
H^{(0)}=\sum _{k=1}^nH_{k,k+1}=
h^{(0)}+h^{(0)*}\,,\,\,[h^{(0)},h^{(0)*}]=0\,.
\end{equation}
It is a consequence of the similar property
for the pair BFKL hamiltonian $H_{12}$ (\ref{H12}) and the
energy $E_{m,\widetilde{m}}$ (\ref{PomE}).

The holomorphic
Hamiltonian in the multi-color QCD can be written as follows (cf. (\ref{H12}))
\begin{equation}
h^{(0)}=\sum _kh^{(0)}_{k,k+1}\,,\,\,h^{(0)}_{12}=\ln
(p_{1}p_{2})+\frac{1}{p_{1}}\,(\ln
\rho _{12})\,p_{1}+
\frac{1}{p_{2}}\,(\ln \rho
_{12})\,p_{2}-2\psi (1)\,,
\label{pairham}
\end{equation}
where $\psi (x)=(\ln \Gamma (x))'$.
As a result, the wave function $\Psi $ has the holomorphic
factorization~\cite{separ}
\begin{equation}
\Psi =\sum _{r,\widetilde{r}}a_{r,\widetilde{r}}\,\Psi ^{r}(\rho _1,...,\rho _n)
\,\Psi ^{\widetilde{r}}(\rho ^*_1,...,\rho ^*_n)\,,
\end{equation}
which in the case of
two-dimensional
conformal field theories is a consequence of the infinite dimensional Virasoro group.
Moreover, the holomorphic hamiltonian $h^{(0)}$ is invariant under the duality
transformation~\cite{dual}
\begin{equation}
p_i\rightarrow \rho _{i,i+1}\rightarrow p_{i+1}\,,
\end{equation}
combined with its transposition.

Further, there are integrals of motion $q_r$
commuting among themselves and with $h^{(0)}$~\cite{int1, int}:
\begin{equation}
q^{(0)}_{r}=\sum_{k_{1}<k_{2}<...<k_{r}}\rho
_{k_{1}k_{2}}\rho
_{k_{2}k_{3}}...\rho
_{k_{2}k_{3}}...\rho
_{k_{r}k_{1}}\,p_{k_{1}}p_{k_{2}}...
p_{k_{r}}\,,\,\,[q_{r},h]=0\,.
\end{equation}
The integrability of the BFKL dynamics in LLA and its relation with the
Baxter XXX-model was established
in ref.~\cite{int}. This remarkable property  is
related also to the fact that $h$ coincides with the local Hamiltonian of an
integrable Heisenberg spin model \cite{Li} (see also~\cite{FK}). Eigenvalues and
eigenfunctions of this hamiltonian were constructed in refs.~\cite{dVL, DKM}
in the framework of the Baxter-Sklyanin approach~\cite{Sklya}.

In the next-to-leading approximation
the integral kernel for the BFKL equation was
calculated in Refs.~\cite{trajN4,FL}.
In QCD the eigenvalue of  the kernel contains
the Kronecker symbols $\delta _{n,0}$ and $\delta _{n,2}$ but in $\mathcal{N}=4$ SYM
it is an analytic function of the conformal spin and
has the property of the maximal
transcendentality~\cite{trajN4,KL}.
This extended $\mathcal{N}=4$ supersymmetric theory appears in the framework of
the AdS/CFT correspondence~\cite{Malda, GKP, W}.
It is important, that the eigenvalues of one-loop anomalous
dimension matrix for twist-2 operators
in $\mathcal{N}=4$ SYM have the maximal transcendentality property and are proportional
to the expression $\psi (1)-\psi (j-1)$,
which is related to the integrability of evolution
equations for quasi-partonic operators in
this model~\cite{L4}. The integrability persists
also for some operators in QCD~\cite{BDMB}. The maximal transcendentality
principle suggested in ref.~\cite{KL} gave a possibility  to extract the
universal anomalous dimension up to three loops in $\mathcal{N}=4$
SYM~\cite{KLV, KLOV} from the corresponding QCD results~\cite{VMV}.
The integrability of the  $\mathcal{N}=4$ model
was demonstrated  for anomalous dimensions of other operators in
higher loops and at large coupling constants~\cite{MZ, BKSZ, BS}.
In particular, the asymptotic
Bethe ansatz allowed to calculate the anomalous dimensions in
four loops~\cite{KLRSV}. This result is in an agreement with the
next-to-leading BFKL predictions
after taking into account the wrapping effects~\cite{BJL}.
The maximal transcendentality was helpful for finding a closed
integral equation for the cusp anomalous
dimension in this model~\cite{ES, BES} with the use of
the 4-loop result~\cite{Bern:2006ew}. The thermodinamic Bethe ansatz
and the approach based on the $Y$-systems allows to calculate the
spectrum of anomalous dimensions in $\mathcal{N}=4$ SYM for an arbitrary
coupling constant~\cite{GKV}.

In recent years a new line of investigations has been started, which
also shows remarkable properties of $\mathcal{N}=4$ SYM: the study of scattering amplitudes.
A few years ago, Bern, Dixon and Smirnov (BDS) suggested
a simple ansatz for the gluon scattering amplitudes in this model~\cite{BDS}.
This ansatz was verified for the elastic amplitude in the strong coupling
regime using the AdS/CFT correspondence~\cite{AldayMalda}. But the BDS
hypothesis does not agree in this regime with the calculation of the multi-particle
amplitude~\cite{Alday:2007he}, leading to the conclusion that a non-vanishing 
remainder function, $R^{(n)}$, has to exist which provides the necessary 
corrections to the BDS amplitudes. The property of the conformal
invariance of the BDS amplitudes in the momentum space was discussed in
ref.~\cite{DHSS}, and the relation with the Wilson loop approach was suggested
in ref.~\cite{Drummond:2007bm} generalizing the results of the strong coupling
calculations of ref.~\cite{AldayMalda}. However, in ref.~\cite{BLS1} it was found 
that the BDS amplitudes $A_n$ for $n\ge 6$
in the multi-Regge kinematics do not have correct analytic properties
compatible with the Steinmann relations~\cite{Steinmann}. It is a consequence of the
fact, that these amplitudes do not contain the Mandelstam cuts~\cite{BLS1}.
The cut contribution was obtained from the BFKL-like
equation for the amplitude with the $
t$-channel exchange of two reggeized gluons 
in the adjoint representation of the gauge group~\cite{BLS1}. This equation
was solved in LLA and the two-loop expression for the 6-point
scattering amplitude in the multi-Regge kinematics was derived~\cite{BLS2}.
The two-loop correction to the remainder function was calculated numerically for some values
of external momenta in an agreement with expectations based on the Wilson loop
approach~\cite{twoloop}. In a recent paper~\cite{Bartels:2010ej}, by solving the set of $Y$-equations,  
also the strong coupling limit of the remainder function has been studied.   

The existence of the Mandelstam cut contribution, found first for the $6$-point amplitude, 
generalizes to multiparticle amplitudes with $n>6$. As an example, 
the $2 \to 2n$ amplitude will contain Mandelstam-cut 
contributions composed of $n$ reggeized $t$-channel gluons. These $n$ gluon $t$-channel 
states can be expressed in terms of solutions of the BKP-like equation in the adjoint 
representation. Most remarkable, in LLA the corresponding Hamiltonian is integrable: 
it  coincides with the local Hamiltonian of an
integrable open Heisenberg spin chain~\cite{Openchain}.    

In this review we present a summary of the Mandelstam cut contributions to 
the inelastic scattering amplitudes in $N=4$ SYM and 
their properties of integrability. We first review the analytic structure of 
$n$-point amplitudes in the multi-Regge kinematics and  describe the main features of the 
Mandelstam cut contributions. In the subsequent section we compare our results in LLA 
with the exact two-loop calculations of Goncharov, Spradlin, Vergu and 
Volovich~(GSVV)~\cite{GSVV} and consider a relation with collinear kinematics. The rest  of the review is devoted 
to the integrability of the BKP Hamiltonian in the adjoint representation.        

\section{The analytic structure of scattering amplitudes in the Regge limit}\label{JB}

Let us begin with a brief summary of the analytic properties of scattering amplitudes 
in the multi-Regge limit.
It is well-known that the  $2 \to 3$ amplitude in the multi-Regge 
kinematics with the exchanged reggeons having definite signatures $\tau_i=\pm 1$ in the crossing channels $t_1$ and $t_2$ can be written as a sum of two terms  
\beq\label{M23}
\frac{M_{2\rightarrow 3}^{pole}}{\Gamma (t_1)\Gamma (t_2)}=
|s_1|^{\omega _{12}}|s|^{\omega _2} \xi_{12} \xi_2 \, \kappa _{12}^{\omega _2}c_1^{12}+
|s_2|^{\omega _{21}}|s|^{\omega _1} \xi_{21} \xi_1 \, \kappa _{12}^{\omega _1}c_2^{12}\,,\,\,
\kappa _{12}=\mathbf{k}_a^2=\frac{s_1s_2}{s}\,,
\eeq
where 
\beq
\xi _1=e^{-i\pi \omega _1}-\tau _1\,,\,\,\xi _2=e^{-i\pi \omega _2}-\tau _2\,,\,\,
\xi _{12}=e^{-i\pi \omega _{12}}+
\tau _1\tau _2\,,\,\,\xi _{21}=e^{-i\pi \omega _{21}}+
\tau _1\tau _2\,.
\eeq
$\Gamma (t_i)$ are the residue functions of the exchanged Regge poles, and $\mathbf{k}_a$ is the 
transverse momentum of the produced particle.
In (\ref{M23}) we assumed that $s, s_1, s_2$ and $\kappa_{12}$ are measured in  some characteristic mass $\mu^2$.

The gluon Regge trajectories in $N=4$ SYM can be written as (see (\ref{reggetraj}))
\beq
\omega _i=\omega (-\mathbf{q}_i^2)= -\frac{\gamma_K}{4}\,\ln \frac{\mathbf{q}_i^2}{\lambda ^2}\,,\,\,
\gamma _K\approx 4a\,,\,\,a=
\frac{g^2\,N_c}{8\pi ^2}\,,\,\,\omega _{12}=\omega _1-\omega _2\,,
\eeq
where $t_i=-\mathbf{q}_i^2$, $\gamma _K$ is the cusp anomalous dimension and $\lambda ^2 \simeq \mu ^2 \exp (1/\epsilon )$
for $D=4-2\epsilon$ with  $\epsilon \rightarrow -0$. The parameter $\lambda ^2$ can be considered as  an effective mass of gluon. 
The real coefficients $c_1^{12},\,c_2^{12}$ in $\mathcal{N}=4$ SYM 
are obtained from the BDS amplitude and  given below~\cite{BLS1}
\beq
c_1^{12}=|\Gamma _{12}|\,\frac{\sin \pi (\omega _1-\omega _a)} {\sin \pi \omega _{12}}\,,
\,\,
c_2^{12}=|\Gamma _{12}|\,\frac{\sin \pi (\omega _2-\omega _a)} {\sin \pi \omega _{21}}\,,
\eeq
where the Reggeon-Reggeon-gluon vertex $\Gamma _{12}$ in the physical region $s,s_1,s_2>0$ is
\beq
\Gamma _{12}(\ln \kappa _{12}-i\pi)=|\Gamma _{12}|\,\exp (i\pi \,\omega _a)\,,\,\,
\omega _a =\frac{\gamma _K}{8}\,
\ln \frac{\mathbf{k}_a^2\lambda ^2}{\mathbf{q}_1^2\mathbf{q}_2^2}\,,
\label{omegaa}
\eeq
\beq
\ln |\Gamma _{12}|=\frac{\gamma _K}{4}\, \left(-\frac{1}{4}\ln ^2\frac{\mathbf{k}_a^2}{\lambda ^2}
-\frac{1}{4}\ln ^2\frac{\mathbf{q}_1^2}{\mathbf{q}_2^2}+
\frac{1}{2}\ln \frac{\mathbf{q}_1^2\mathbf{q}_2^2}{\lambda ^4}\ln \frac{\mathbf{k}_a^2}{\mu ^2}
+\frac{5}{4}\zeta _2\right)\,.
\eeq

The amplitude  representation in (\ref{M23}) is compatible with the
Steinmann relations~\cite{Steinmann} forbidding the simultaneous singularities in the overlapping channels $s_1$ and $s_2$.
 The expression in (\ref{M23})  can be rewritten in the factorized form
\beq
\frac{M^{\tau _1\tau _2}_{2\rightarrow 3}}{\Gamma (t_1)\Gamma (t_2)}=
|s_1|^{\omega _{1}}\xi _1\,V^{\tau _1\tau _2}
\,|s_2|^{\omega _2}\xi _2\,,\,\,V^{\tau _1\tau _2}=\frac{\xi _{12}}{\xi _1}\,c_1^{12}+
\frac{\xi _{21}}{\xi _2}\,c_2^{12}\,.
\label{sign23}
\eeq

The BDS-amplitude which holds in the planar approximation can also be written 
as a sum of two terms:
\beq
\frac{M_{2\rightarrow 3}^{BDS}}{\Gamma (t_1)\Gamma (t_2)}=
(-s_1)^{\omega _{12}}(-s\kappa _{12})^{\omega _2}c_1^{12}+
(-s_2)^{\omega _{21}}(-s\kappa _{12})^{\omega _1}c_2^{12}\,,
\eeq
where we put the normalization point $\mu ^2$ in the Regge factors equal to unity. 

\begin{figure}[htbp]
	\begin{center}
		\epsfig{figure=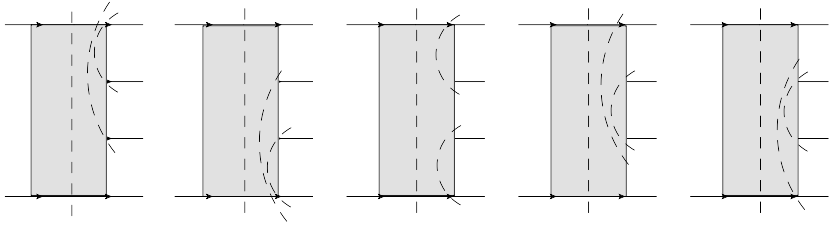,width=110mm}
	\end{center}
	\caption{ The analytic representation of the $2 \to 4$ scattering amplitude.
The dashed lines denote  possible energy discontinuities.}
	\label{fig:2to4-fiveterms}
\end{figure}

For the $2\to 4$ amplitude in the multi-Regge kinematics the situation is more 
complicated. In agreement with the Steinmann relations, the Regge-pole scattering amplitude    
can, again, be written as a sum of 5 terms as illustrated in Fig.~\ref{fig:2to4-fiveterms}.

For the signatured amplitude, this 
representation   is equivalent to the factorized form:  
\beq
\frac{M^{\tau _1\tau _2\tau _3}_{2\rightarrow 4}}{\Gamma (t_1)\Gamma (t_3)}=
|s_1|^{\omega _{1}}\xi _1\,V^{\tau _1\tau _2}
\,|s_2|^{\omega _2}\xi _2\,V^{\tau _2\tau _3}\,|s_3|^{\omega _3}\xi _3\,,
\label{sign24}
\eeq
where $V^{\tau _2\tau _3}$ is obtained from $V^{\tau _1\tau _2}$ (\ref{sign23})
with the corresponding substitutions
\beq
V^{\tau _2\tau _3}=\frac{\xi _{23}}{\xi _2}\,c_1^{23}+
\frac{\xi _{32}}{\xi _3}\,c_2^{23}\,.
\eeq
For the second produced gluon with the transverse momentum $k_b$ the coefficients
$c^{23}$ and phase $\omega _b$ read
\beqn
c_1^{23}=|\Gamma _{23}|\,\frac{\sin \pi (\omega _2-\omega _{b})} {\sin \pi \omega _{23}},\;
c_2^{23}=|\Gamma _{23}|\,\frac{\sin \pi (\omega _3-\omega _{b})} {\sin \pi \omega _{32}},\;
\omega _{b}= \frac{\gamma _K}{8}\ln \frac{\mathbf{k}_b^2\lambda ^2}{\mathbf{q}_2^2\mathbf{q}_3^2},\,
\mathbf{k}_b^2=\left|\frac{s_2s_3}{s_{123}}\right|.
\label{omegab}
\eeqn
In the planar approximation one expects that the scattering amplitude, 
$M_{2\rightarrow 4}^{pole}$, in accordance with the Steinmann relations~\cite{Steinmann}, has the 
form~\cite{BLS1}:
\[
\frac{M_{2\rightarrow 4}^{pole}}{\Gamma (t_1)\Gamma (t_3)}=(-s_1)^{\omega _{12}}\,
(-s_{012}\kappa _{12})^{\omega _{23}}\,(-s\kappa _{12}\kappa _{23})^{\omega _3}\,
c_1^{12}\,c_1^{23}
\]
\[
+(-s_3)^{\omega _{32}}
(-s_{123}\kappa _{23})^{\omega _{21}}\,(-s\kappa _{12}\kappa _{23})^{\omega _1}\,c_2^{12}\,
c_2^{23}
+(-s\kappa _{12}\kappa _{23})^{\omega _2}\,(-s_1)^{\omega _{12}}\,(-s_3)^{\omega _{32}}\,
c_1^{12}\,c_2^{23}
\]
 \[
 +(-s_2)^{\omega _{21}}
 (-s_{012}\kappa _{12})^{\omega _{13}}\,(-s\kappa _{12}\kappa _{23})^{\omega _3}\,
 \frac{\sin \pi \omega _1}{\sin \pi \omega _2}\,
 \frac{\sin \pi \omega _{23}}{\sin \pi \omega _{13}}\,c_2^{12}\,c_1^{23}
 \]
\beq
 +(-s_2)^{\omega _{23}}
 (-s_{123}\kappa _{23})^{\omega _{31}}\,(-s\kappa _{12}\kappa _{23})^{\omega _1}\,
 \frac{\sin \pi \omega _3}{\sin \pi \omega _2}\,
 \frac{\sin \pi \omega _{21}}{\sin \pi \omega _{31}}\,c_2^{12}\,c_1^{23}\,.
\label{Stau24}
 \eeq

A closer look at the last two terms shows that a model with Regge-poles only 
exhibits unphysical poles, indicating that a pure Regge model maybe   
incompatible with the correct analytic structure of multiparticle amplitudes in the 
multi-Regge kinematics. In fact, the LLA analysis of N=4 SYM gauge theory has shown that, 
in addition to the gluon Regge pole, there exists also a Mandelstam cut in the complex angular momentum plane, 
which removes this inconsistency.
 \begin{figure}[htbp]
	\begin{center}
		\epsfig{figure=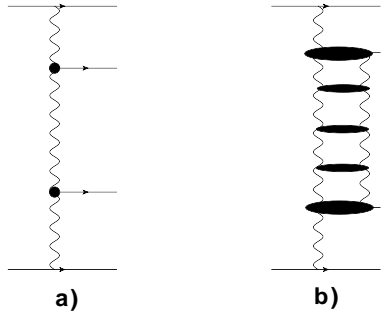,width=50mm}
	\end{center}
	\caption{ Contributions of $\mathbf{a)}$ the Regge poles  and $\mathbf{b)}$ the Mandelstam cut   to the $2 \to 4$ scattering amplitude in the  $t_2$-channel. The wavy lines represent reggeized gluons. The Mandelstam cut appears as a bound state of two reggeized gluons. }
	\label{fig:2to4reggecut}
\end{figure}
The Mandelstam cut  appears in the angular momentum plane of the $t_2$ channel and describes a bound state of two or more reggeized gluons as depicted in Fig.~\ref{fig:2to4reggecut}.
In the planar approximation, it shows up in the special physical kinematic 
regions where the invariants in the direct channels
have the following signs $s,s_2>0;\,s_1,s_3,s_{012},s_{123}<0$ or
$s,s_1,s_2,s_3<0;\,s_{012},s_{123}>0$~\cite{BLS1}, and it is not visible in the
physical kinematic region where all energies are positive.  In the following these special physical regions will be named "Mandelstam regions".

The $2 \to 4$ amplitude in the multi-Regge kinematics can be written as a sum
of the Regge pole and Mandelstam cut contributions~\cite{BLS1}
\beq\label{PolCut}
M_{2\rightarrow 4}= M_{2\rightarrow 4}^{pole}+M_{2\rightarrow 4}^{cut}\,,
\eeq
where $M_{2\rightarrow 4}^{cut}$ is 
a generalization of two last terms in (\ref{Stau24}),  and  it is non-zero only in the two kinematic regions
restricted by the inequalities
$s,s_2>0; \,s_1,s_3,s_{012},s_{123}<0$ and $s,s_1,s_2,s_3<0; \,s_{012},s_{123}>0$.  There is some freedom in redistributing terms between the Regge pole $M_{2\rightarrow 4}^{pole}$ and the Mandelstam cut $M_{2\rightarrow 4}^{cut}$ contributions. Using this fact and representation (\ref{Stau24}) one can write in the region $s,s_2>0; \,s_1,s_3,s_{012},s_{123}<0$~\cite{LipDisp}
\beq
\frac{M_{2\rightarrow 4}^{pole}}{|s_1|^{\omega _{1}}|s_2|^{\omega _{2}}|s_3|^{\omega _{3}}
  |\Gamma _{12}||\Gamma _{23}|
  \,\Gamma (t_1)\Gamma (t_3)}=e^{-i\pi \omega _2}\cos \pi \omega _{ab}\,
  \label{pole2}
\eeq
and
\beq
\frac{M_{2\rightarrow 4}^{cut}}{|s_1|^{\omega _{1}}|s_2|^{\omega _{2}}|s_3|^{\omega _{3}}
  |\Gamma _{12}||\Gamma _{23}|
  \,\Gamma (t_1)\Gamma (t_3)}= i\,e^{-i\pi \omega _2}  \,
\int _{-i\infty}^{i \infty}
\frac{d\omega _{2'}}{2\pi i}\,f(\omega _{2'})\,e^{- i\pi \omega _{2'}}\,
|s_2|^{\omega _{2'}}\, ,
\label{cut}
\eeq
where $\omega _{ab}$ is obtained from   (\ref{omegaa}) and (\ref{omegab}) and reads
\beqn\label{omegaabJB}
\omega _{ab}=\frac{\gamma _K}{8}\,\ln \frac{\mathbf{k}^2_a\mathbf{q}^2_3}{\mathbf{k}^2_b \mathbf{q}^2_1}.
\eeqn

In the other physical region, where $s,s_1,s_2,s_3<0; \,s_{012},s_{123}>0$ (corresponding to the physical channel for the $3 \to 3$ transition) we find~\cite{LipDisp}
\beq\label{pole3}
\frac{M_{2\rightarrow 4}^{pole}}{|s_1|^{\omega _{1}}|s_2|^{\omega _{2}}|s_3|^{\omega _{3}}
  |\Gamma _{12}||\Gamma _{23}|
  \,\Gamma (t_1)\Gamma (t_3)}=\cos \pi \omega _{ab}\,
\eeq
and
\beq\label{cut3}
\frac{M_{2\rightarrow 4}^{cut}}{|s_1|^{\omega _{1}}|s_2|^{\omega _{2}}|s_3|^{\omega _{3}}
  |\Gamma _{12}||\Gamma _{23}|
  \,\Gamma (t_1)\Gamma (t_3)}= -i\,
\int _{-i\infty}^{i \infty}
\frac{d\omega _{2'}}{2\pi i}\,f(\omega _{2'})\,
|s_2|^{\omega _{2'}}\,.
\eeq
The function $f(\omega _{2'})$ is pure real and  denotes the partial wave in the complex angular momentum plane.

\begin{figure}[htbp]
	\begin{center}
		\epsfig{figure=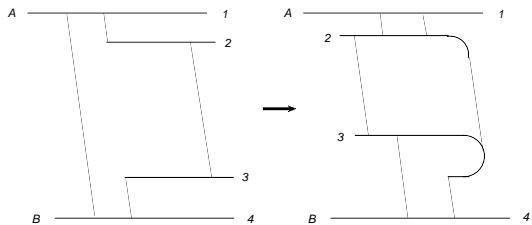,width=70mm}
	\end{center}
	\caption{ Twisting a planar $2 \to 4$ amplitude. }
	\label{fig:2to4mixed}
\end{figure}

The origin of  
this 'restricted' appearance of the  Mandelstam cut contribution  $M_{2\rightarrow 4}^{cut}$ can be traced back to 
Mandelstam's argument for the existence of the Regge  cuts (Fig.~\ref{fig:mandelstam}): 
\begin{figure}[htbp]
	\begin{center}
		\epsfig{figure=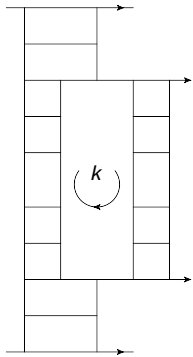,width=25mm}
	\end{center}
	\caption{The diagrammatic structure of the Mandelstam cut. }
	\label{fig:mandelstam}
\end{figure}
If we put for the reggeon momentum $k=\alpha p_A + \beta p_B + k_{\perp}$, it is easy to see that, 
in the planar (large-$N_c$) limit with all energies being positive, the 
integrations over $\alpha$ and $\beta$ have singularities only in the upper half planes and 
lead to the absence of the Mandelstam cut contribution. However, if by pulling the produced particles to the left we 'twist' 
the reggeons (ladders) in the $t_1$ and $t_3$ channels (Fig.~\ref{fig:2to4mixed}), there will be 
singularities on both sides of the real $\alpha$ and $\beta$ axis, 
and the Mandelstam singularity remains. Note that, despite this 'twisting', the amplitude is still 
planar.
Returning to the sum of the five contributions in Fig.~\ref{fig:2to4-fiveterms}, it 
can be shown that 
the Mandelstam cut contribution should be present only in the last two terms: taking into account the phase 
structure one finds that, in 
the physical kinematic region, where all energies are positive, the cut cancels in 
the sum of the two terms. This cancellation does not work, if we are in the 
'mixed' regions $s,s_2>0;\,s_1,s_3,s_{012},s_{123}<0$ or $s,s_1,s_2,s_3<0;\,s_{012},s_{123}>0$. 

Turning now to the BDS amplitudes, this Mandelstam cut contribution is missing and must therefore 
be contained in the remainder function, $R_{ 2\to 4}$. It is believed that 
the full MHV amplitude in the planar (large-$N_c$) approximation can be written in the 
factorized form:
\beq
M_{2 \to 4} = M^{BDS}_{2 \to 4}\,R_{ 2\to 4}.
\eeq
Indeed, in~\cite{BLS2} 
it was shown that in the region $s,s_2>0;\,s_1,s_3,s_{012},s_{123}<0$ the correct form
of the LLA $2 \to 4$ scattering amplitude is       
\beqn
M^{LLA}_{2\rightarrow 4}=M^{BDS}_{2\rightarrow 4}\,(1+i\Delta^{LLA} _{2\rightarrow 4}),
\label{corLLAJB}
\eeqn
where $M^{BDS}_{2\rightarrow 4}$ is the BDS amplitude~\cite{BDS} and
\beq
\Delta^{LLA} _{2\rightarrow 4}=\frac{a}{2}\, \sum _{n=-\infty}^\infty (-1)^n
\int _{-\infty}^\infty \frac{d\nu }{\nu ^2+\frac{n^2}{4}}\,
\left(\frac{q_3^*k^*_a}{k^*_bq_1^*}\right)^{i\nu -\frac{n}{2}}\,
\left(\frac{q_3k_a}{k_bq_1}\right)^{i\nu +\frac{n}{2}}\,
\left(s_2^ {\omega (\nu , n)}-1\right)\,.
\label{LLAJB} 
\eeq
Here $k_a,k_b$ are the complex transverse components of the produced gluon momenta,
$q_1,q_2,q_3$ are the momenta of reggeons in the corresponding
crossing channels, and
\beq
\omega (\nu , n)=4a\,
\Re \left(2\psi (1)-\psi (1+i\nu +\frac{n}{2})-\psi (1+i\nu -\frac{n}{2})\right)
\label{eigen}
\eeq
is the eigenvalue of the  BFKL Hamiltonian in the adjoint representation. 
The correction $\Delta^{LLA}_{2 \to 4}$ is M\"{o}bius invariant in the momentum space,
and it is important to note that it can be written in terms of the 
four-dimensional anharmonic ratios~\cite{BLS2} in an accordance
with the results of refs.~\cite{DHSS}. Thus it can be viewed as part 
of the remainder function, $R_{ 2\to 4}$, which is expected to depend only on the 
three anharmonic ratios $u_i$, $i=1,2,3$.   
In section 3 we will come back for a closer look at this expression.

\begin{figure}[htbp]
	\begin{center}
		\epsfig{figure=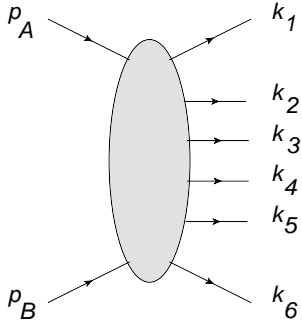,width=40mm}
	\end{center}
	\caption{ The $2 \to 6$ scattering amplitude. }
	\label{fig:2to6}
\end{figure}

\begin{figure}[htbp]
	\begin{center}
		\epsfig{figure=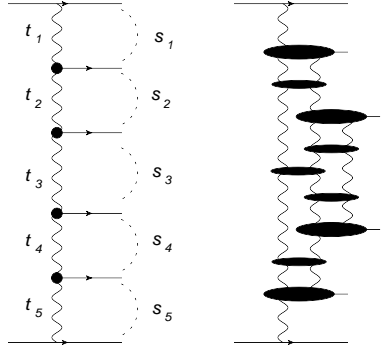,width=50mm}
	\end{center}
	\caption{ The Regge pole and two- and three-gluon cuts in the $2 \to 6$ amplitude. }
	\label{fig:2to6reggecut}
\end{figure}

\begin{figure}[htbp]
	\begin{center}
		\epsfig{figure=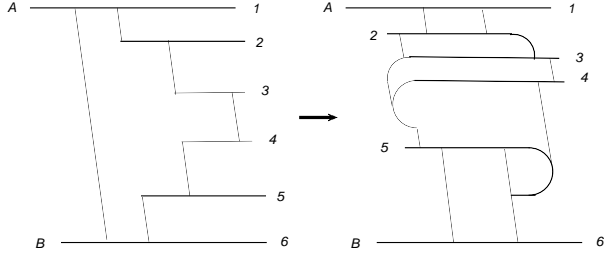,width=80mm}
	\end{center}
	\caption{ The  'double twisting' of the $2 \to 6$ amplitude. }
	\label{fig:2to6mixed}
\end{figure}

As we have already mentioned in the introduction, this  Mandelstam cut structure of six-point amplitudes 
in the multi-Regge kinematics can be generalized to a larger number of external 
legs, $n>6$. This generalization exhibits the remarkable feature of integrability~\cite{Openchain}. 
As an example, let us consider       
the eight-point amplitude $M_{2 \to 6}$ in multi-Regge kinematics (Fig.~\ref{fig:2to6}).
Again, the amplitude can be written as a sum of terms which are compatible 
with the Steinmann relations. The number of terms is already 42 \footnote{
for the $2 \to n$ amplitude, the number of terms, $N_n$, obeys the 
recursion relation $N_n = \sum_{k=1}^{n-1}N_k N_{n-k}$ with $N_1=N_2=1$.} 
and will not be discussed here in further detail. 
We only mention that, for the fully-signatured amplitude, the Regge-pole contribution can also be 
rewritten in the factorized representation (cf.(\ref{sign24})):  
\beq
\frac{M^{\tau _1\tau _2\tau _3\tau_4}_{2\rightarrow 6}}{\Gamma (t_1)\Gamma (t_4)}=
|s_1|^{\omega _{1}}\xi _1\,V^{\tau _1\tau _2}
\,|s_2|^{\omega _2}\xi _2\,V^{\tau _2\tau _3}\,|s_3|^{\omega _3}\xi _3\,
V^{\tau _3\tau _4}\,|s_4|^{\omega _4}\xi _4\,.
\label{sign26}
\eeq      
As it was already the case for the $2 \to 4$ scattering amplitude, there 
exist Mandelstam cut contributions which appear only in special physical regions.   
The most interesting one is illustrated in Fig.~\ref{fig:2to6reggecut}. 
This Regge-cut piece 
belongs to singularities in the angular momentum variables $j_2$, $j_3$, and $j_4$  of the 
$t_2$, $t_3$, and $t_4$ channels, resp. Using the Mandelstam 
argument given above, it is easy to see that it appears, for example, in the 
physical region $ s_{3}>0$, $s_{2345}>0$, $s>0$, $s_1<0$, $s_2<0$, 
$s_4<0$, $s_5<0$ ~\cite{Openchain}. This region is obtained by 'double twisting' of the planar amplitude and is 
further illustrated in Fig.~\ref{fig:2to6mixed}.
\newpage

The detailed form of this contribution will be given in a forthcoming paper. 
Here we only mention that the singularities in $j_2$ and $j_4$ are described 
by the  BFKL Hamiltonian of the two reggeon state in the adjoint representation, whereas the $j_3$ channel is governed 
by the spectrum of the BKP Hamiltonian of three reggeized gluons, projected 
on the   adjoint representation. In section 3 we will give a more detailed discussion:
in particular, it will be shown that this Hamiltonian is integrable and belongs to 
an open spin chain.

\section{BFKL approach and MHV amplitudes}\label{AP}

In this section we discuss six-particle amplitudes in the multi-Regge kinematics 
in some of  the Mandelstam regions. We compare results obtained in the BFKL approach 
with those calculated using Wilson Loop/Scattering Amplitude duality at two loops.
In particular, the analysis of the two-loop result allows to obtain the impact factor 
for the Mandelstam cut contribution beyond the LLA. 
We also discuss briefly the collinear limit and write an explicit analytic form of the all-loop remainder function in the Double Leading  Logarithmic Approximation~(DLLA).   The section consists of two parts devoted to  $2 \to 4 $  and  $3 \to 3 $  amplitudes.

\subsection{$2 \to 4 $  amplitude}\label{sec:regge}
\begin{figure}[htbp]
	\begin{center}
		\epsfig{figure=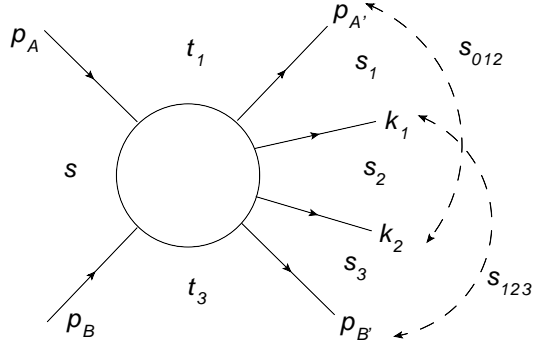,width=70mm}
	\end{center}
	\caption{ The $2 \to 4$ gluon scattering amplitude. }
	\label{fig:2to4ope}
\end{figure}

The six-particle scattering amplitude  corresponds to two physical processes, namely to $2 \to 4$ and  $3 \to 3$ scattering. Firstly we discuss the $2 \to 4$ planar MHV amplitude illustrated in Fig.~\ref{fig:2to4ope} and review the main result of the BFKL analysis applied to this case. The corresponding Mandelstam variables  are defined as  $s=(p_A+p_B)^2, \;
s_1=(p_{A'}+k_1)^2, \;  s_{2}=(k_1+k_2)^2, \; s_3=(p_{B'}+k_2)^2,  \;s_{012}=(p_{A'}+k_1+k_2)^2, \;  s_{123}=(p_{B'}+k_1+k_2)^2, \; t_{1}=(p_A-p_{A'})^2, \; t_{2}=(p_A-p_{A'}-k_1)^2 $, $t_3=(p_{B}-p_{B'})^2$ and the dual conformal cross ratios are given by
\beqn\label{crossinv}
u_1=\frac{s s_{2}}{s_{012}\; s_{123}},\; u_2=\frac{s_{1}t_{3}}{s_{012}\; t_2}, \; u_3=\frac{s_{3}t_{1}}{s_{123}  t_2}.
\eeqn

\begin{figure}[htbp]
	\begin{center}
		\epsfig{figure=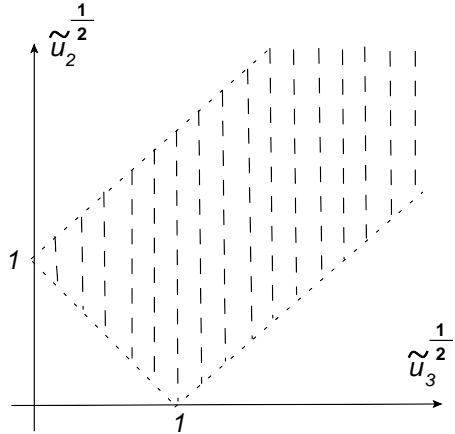,width=60mm}
	\end{center}
	\caption{ The physical values of $\sqrt{\tilde{u}_2}$ and $\sqrt{\tilde{u}_3}$ lie in the shaded semi-infinite strip. }
	\label{fig:region}
\end{figure}

The multi-Regge kinematics, where
\beqn\label{MRK}
-s \gg -s_{012}, -s_{123} \gg -s_1, -s_2, -s_3 \gg -t_1, -t_2, -t_3 > 0
\eeqn
 implies
\beqn\label{multicross}
1-u_1 \to +0,\;\; u_2 \to +0, \;\; u_3 \to +0, \;\; \frac{u_2}{1-u_1} \simeq \mathcal{O}(1), \;\; \frac{u_3}{1-u_1} \simeq \mathcal{O}(1),
\eeqn
which suggests that in this kinematics the convenient variables for the remainder function are $1-u_1$ and the reduced cross ratios  defined by
\beqn\label{redcross}
 \tilde{u}_2=\frac{u_2}{1-u_1}, \;\;\tilde{u}_3=\frac{u_3}{1-u_1}.
\eeqn
In the Regge limit they can be expressed through $s_2$ and the transverse momenta
\beqn\label{redcrossperp}
1-u_1\simeq \frac{(\mathbf{k}_1+\mathbf{k}_2)^2}{s_2}, \;\; \tilde{u}_2\simeq\frac{\mathbf{k}_1^2 \; \mathbf{q}^2_3}{(\mathbf{k}_1+\mathbf{k}_2)^2 \; \mathbf{q}_2^2}, \;\; \tilde{u}_3 \simeq\frac{\mathbf{k}_2^2 \; \mathbf{q}^2_1}{(\mathbf{k}_1+\mathbf{k}_2)^2 \; \mathbf{q}_2^2}.
\eeqn
Note that $\tilde{u}_2$ and $\tilde{u}_3$ are rational functions of cross ratios in four dimensions, but in the Regge limit they are simple cross ratios in the two-dimensional transverse space as one can see from (\ref{redcrossperp}). The region of possible values of $\tilde{u}_2$ and $\tilde{u}_3$ that correspond to physical momenta is depicted in Fig.~\ref{fig:region} as a semi-infinite strip~\cite{LP1,LP2}. 

\begin{figure}[htbp]
	\begin{center}
		\epsfig{figure=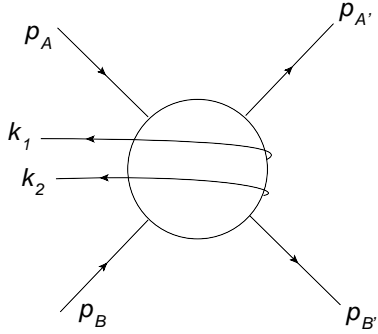,width=50mm}
	\end{center}
	\caption{ The Mandelstam channel of the $2 \to 4$ gluon planar scattering amplitude. }
	\label{fig:2to4opeINV}
\end{figure}

In the ``Euclidean'' kinematics, where all  invariants are negative  and thus all $u_i$ are positive, the remainder function vanishes asymptotically  as follows from the analysis presented in refs.~\cite{BLS1,BLS2}.
However, these studies also show that this is not the case in a slightly different physical region, where one  or more dual conformal cross ratios possess a phase.  This happens when some energy invariants change the sign and here we consider one of such  regions of the  $2 \to 4$ scattering amplitude having
\beqn\label{u1cont}
u_1=|u_1| e^{-i2\pi}, \; u_2 \;\text{and} \; u_3  \;\text{are fixed and positive}.
\eeqn
It corresponds to  the physical region mentioned before~(the Mandelstam region), where
\beqn\label{mandels}
s\; ,s_2> 0; \;\;\;s_{1},\;s_{3},\;s_{012},\;s_{123}<0
\eeqn
as illustrated in Figs.~\ref{fig:2to4mixed} and \ref{fig:2to4opeINV}. The $t$-variables are all  negative in the physical regions under consideration for the $2 \to 4$ scattering amplitude in the Regge kinematics.  It is worth emphasizing that the scattering amplitude in Fig.~\ref{fig:2to4opeINV} is still planar, but the produced particles have reversed momenta $k_1$ and $k_2$ with a negative energy components.

In the Mandelstam channel  the remainder function grows with energy $s_2$ and was first calculated using the BFKL approach by two of the authors in collaboration with A.~Sabio Vera in ref.~\cite{BLS2}. The BFKL approach, based   on the analyticity and  unitarity was developed more than thirty years ago~\cite{BFKL}. In this approach  one sums the contributions from the Feynman diagrams, which are enhanced by the logarithms of the energy~($1-u_1\simeq (\mathbf{k}_1+\mathbf{k}_2)^2/s_2$ in our case). The Leading Logarithmic Approximation~(LLA) allows to write an integral representation of the remainder function $R$ to any order of the parameter $g^2  \ln s_2 $. As it was already discussed in the previous section~(see (\ref{corLLAJB})), the amplitude  in this Mandelstam channel  is given by~\cite{BLS2}
\beq
M_{2\rightarrow 4}=M^{BDS}_{2\rightarrow 4}\,R_{2 \to 4} =M^{BDS}_{2\rightarrow 4}\,(1+i\Delta _{2\rightarrow 4}),
\label{corLLA}
\eeq
where $M^{BDS}_{2\rightarrow 4}$ is the BDS expression~\cite{BDS} and the correction $\Delta _{2\rightarrow 4}$ was calculated
in all orders with a leading logarithmic accuracy using the solution to the  BFKL equation in the adjoint representation. The all-order LLA expression  for $\Delta_{2\to 4}$  was given in  (\ref{LLAJB})
\beqn\label{LLA}
  \Delta^{LLA} _{2\rightarrow 4}
\simeq \frac{a}{2}\, \sum _{n=-\infty}^\infty (-1)^n
\int _{-\infty}^\infty \frac{d\nu }{\nu ^2+\frac{n^2}{4}}\,
\left(w^*\right)^{i\nu -\frac{n}{2}}\,
\left(w\right)^{i\nu +\frac{n}{2}}\,
\left((1-u_1)^ {-\omega (\nu , n)}-1\right).
\eeqn
Here $k_1,k_2$ are complex transverse components of the gluon momenta,
$q_1,q_2,q_3$ are the corresponding  momenta of reggeons in the
crossing channels. It is convenient to define holomorphic and antiholomorphic variables  in the transverse space as
\beqn
w =\frac{q_3k_1}{k_2q_1}, \;\; w^* =\frac{q^*_3k^*_1}{k^*_2q^*_1}
\eeqn
related to the reduced  cross ratios of (\ref{redcross}) by
\beqn\label{wdef}
 |w|^2=\frac{\tilde{u}_2}{\tilde{u}_3}=\frac{u_2}{u_3}, \;\; w=|w| e^{i(\phi_2-\phi_3)},\;\; \cos (\phi_2-\phi_3)=\frac{1-\tilde{u}_2-\tilde{u}_3}{2 \sqrt{\tilde{u}_2 \tilde{u}_3}}=\frac{1-u_1-u_2-u_3}{2 \sqrt{u_2 u_3}}.
\eeqn
 The energy behavior of the remainder function is determined by the Mandelstam cut intercept
\beq\label{eigen}
\omega (\nu , n)=-a E_{\nu,n},
\eeq
where $a$  and $E_{\nu,n}$  are the  perturbation theory parameter and  the 
eigenvalue of the BFKL Hamiltonian in the adjoint representation  given by
\beqn\label{Enun}
a=
\frac{\alpha_s N_c}{2\pi}, \;\;\;E_{\nu,n}=-\frac{1}{2}\frac{|n|}{\nu^2+\frac{n^2}{4}}+\psi\left(1+i\nu+\frac{|n|}{2}\right)+\psi\left(1-i\nu+\frac{|n|}{2}\right)-2\psi(1).
\eeqn
Here $\psi(z)=\Gamma'(z)/\Gamma(z)$ and  $\gamma=-\psi(1)$ is the Euler constant.
The two loop LLA expression for the remainder function in the BFKL approach was first found  from (\ref{corLLA}) and (\ref{LLA}) in ref.~\cite{BLS2} and it reads
\beqn\label{R2LLABFKL}
R^{(2)\; LLA}_{2 \to 4 }=\frac{i\pi}{2} \ln(1-u_1) \ln \tilde{u}_2\ln \tilde{u}_3= \frac{i\pi}{2} \ln(1-u_1) \ln\left|1+w\right|^2 \ln\left|1+\frac{1}{w}\right|^2.
\eeqn
The remainder function in (\ref{R2LLABFKL}) is pure imaginary and symmetric under $w \to 1/w$ transformation, which corresponds to the target-projectile symmetry $p_{A} \leftrightarrow p_{B},\; p_{A'} \leftrightarrow p_{B'}$ and $k_{1} \leftrightarrow k_{2}$ in accordance with (\ref{LLA}).

This result was shown by Schabinger~\cite{Schabinger} to agree numerically with the analytic continuation of the expression for the two-loop remainder function found by Drummond, Henn, Korchemsky and Sokatchev~\cite{hexagon} from  Wilson Loop/Scattering Amplitude duality.
  A rather complicated expression of ref.~\cite{hexagon} was   largely simplified by Del~Duca,~Duhr~and~Smirnov~\cite{DelDuca:2009au,DelDuca:2010zg} and then by Goncharov, Spradlin, Vergu and Volovich~(GSVV)~\cite{GSVV}.
 The prediction in (\ref{R2LLABFKL}) was   analytically confirmed by two of the authors~\cite{LP1} performing the analytic continuation of the GSVV expression for the remainder function at two loops.  The analytic continuation allowed also to extract the next-to-leading logarithmic~(NLLA)  contribution, not yet available from the BFKL approach
\begin{eqnarray}\label{R2NLLABFKL}
 &&R^{(2)\;NLLA}_{2 \to 4 }=
 \frac{i\pi}{2} \ln |w|^2 \ln^2|1+w|^2-\frac{i\pi}{3}\ln^3 |1+w|^2+ i\pi \ln |w|^2 \left( \text{Li}_2 (-w) +\text{Li}_2 (-w^*)\right)
\nonumber \\
&&-i2\pi  \left( \text{Li}_3 (-w) +\text{Li}_3 (-w^*)\right).
\end{eqnarray}
The NLLA remainder function  in (\ref{R2LLABFKL}) is also pure imaginary and symmetric under  
$w \to 1/w$ transformation.  Both of the contributions are pure imaginary due to a 
cancellation of the real part coming from the Mandelstam cut, Regge pole and a phase 
present in the BDS amplitude as was shown by one of the authors~\cite{LipDisp}. 
Starting at three loops the cancellation does not happen anymore and the real part gives a 
non-vanishing contribution at the next-to-leading level. The analysis of ref.~\cite{LipDisp} 
based on analyticity  and  other general properties of the scattering amplitudes resulted in a 
formulation of the dispersion-like relation for the real and imaginary parts of the remainder 
function in the Regge kinematics  in this Mandelstam region 
\beqn\label{disp}
R_{2 \to 4 } \,e^{i\pi \,\delta}=\cos \pi \omega _{ab}+i\int _{-i\infty}^{i \infty}\frac{d\omega}{2\pi i}\,
f(\omega )\,e^{-i\pi \omega}\,(1-u_1)^{-\omega}\,,
\eeqn
where the first term in RHS corresponds to the contribution of the Regge pole~(see~(\ref{pole2})).
This term  as well as the phase $\delta$ in LHS of (\ref{disp}) are obtained directly from the BDS formula~(see~also~(\ref{omegaabJB}))
\beqn\label{deltaomega}
\delta =\frac{\gamma _K}{8}\,\ln (\tilde{u}_2 \tilde{u}_3)=\frac{\gamma _K}{8} \ln \frac{|w|^2}{|1+w|^4}\,,\,\,
\omega _{ab}
=\frac{\gamma _K}{8}\,\ln \frac{\tilde{u}_2}{\tilde{u}_3}=\frac{\gamma _K}{8} \ln |w|^2.\,
\eeqn
The second terms in RHS of (\ref{disp})  stands for the contribution of the Mandelstam 
cut~(see~(\ref{cut})). The coefficient  $\gamma _K \simeq 4 a$ is the cusp anomalous 
dimension known to an arbitrary  order of the perturbation theory. 
The only unknown piece in (\ref{disp}) is the real function $f(\omega)$, 
which contains the Mandelstam cut in $\omega$, depends only on the transverse particle momenta 
and has no energy dependence. In the leading logarithmic approximation $f(\omega)$ is given by
\beqn\label{fomegaLLA}
&&f^{LLA}(\omega)=\frac{a}{2}\sum_{n=-\infty}^{\infty} \int_{-\infty}^{\infty} d \nu \frac{1}{\omega-\omega(\nu,n)} \frac{(-1)^n}{\nu^2+\frac{n^2}{4}}\left(w^* \right)^{i\nu -\frac{n}{2}}\,
\left(w\right)^{i\nu +\frac{n}{2}},
\eeqn
where $\omega(\nu,n)$ is defined in (\ref{eigen}).

The dispersion-like relation in (\ref{disp})  was used~\cite{LP2} for calculating the three loop contributions to $R^{(3)}_{2 \to 4}$~(leading imaginary and the sub-leading real terms) in the multi-Regge kinematics
\beqn\label{R3LLABFKL}
&&  R_{2 \to 4 }^{(3)\;LLA}=i\Delta^{(3)} _{2\rightarrow 4}/a^3= \frac{i\pi}{4} \ln^2(1-u_1)\left(
 \ln|w|^2\ln^2|1+w|^2-\frac{2}{3}\ln^3|1+w|^2\right. \hspace{1cm}\;\;\;
\\
&&\left.-\frac{1}{4}\ln^2|w|^2 \ln|1+w|^2+\frac{1}{2} \ln|w|^2 \left(\text{Li}_2(-w)+\text{Li}_2(-w^*)\right)
 - \text{Li}_3(-w)-\text{Li}_3(-w^*)\right)  \nonumber
\eeqn
and
\beqn\label{R3NLLABFKL}
&&  \Re\left(R^{(3)\;NLLA}_{2 \to 4 }\right)= \frac{\pi^2}{4} \ln (1-u_1)\left(
 \ln|w|^2\ln^2|1+w|^2-\frac{2}{3}\ln^3|1+w|^2\right. \hspace{1cm}\;\;\;
\\
&&\left.-\frac{1}{2}\ln^2|w|^2 \ln|1+w|^2- \ln|w|^2 \left(\text{Li}_2(-w)+\text{Li}_2(-w^*)\right)
 +2 \text{Li}_3(-w)+2\text{Li}_3(-w^*)\right). \nonumber
 \eeqn
As in the two loop case,  both (\ref{R3LLABFKL})  and (\ref{R3NLLABFKL}) are symmetric 
under $w \to 1/w$ transformation, which is obvious from (\ref{LLA}) and corresponds to 
the target-projectile symmetry of the scattering amplitude. The  corrections, subleading 
in the logarithm of the energy,  are not captured by (\ref{LLA}) and require some knowledge 
of the next-to-leading impact factor and the  eigenvalue of the BFKL Kernel in the adjoint representation. 

While the latter is still to be found from the  next-to-leading BFKL equation constructed by Fadin 
and Fiore~\cite{Fadin:2004zq,Fadin:2005zj}, the correction to the impact factor was 
obtained in ref.~\cite{LP2} extracting it from (\ref{R2NLLABFKL}).
\begin{figure}[htbp]
	\begin{center}
		\epsfig{figure=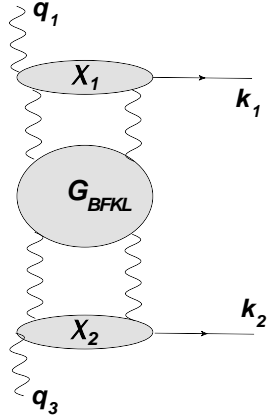,width=35mm}
	\end{center}
	\caption{ A graphic representation of the expression in (\ref{LLA}). Two impact factors $\chi_1$ and $\chi_2$ are convoluted with the propagator of the BFKL state $G_{BFKL}$. }
	\label{fig:chiGchi}
\end{figure}
The integrals in (\ref{LLA}) and (\ref{disp}) come as a convolution of the propagator of the BFKL state $G_{BFKL}$ and two impact factors $\chi_1$ and $\chi_2$ as shown in Fig.~\ref{fig:chiGchi}. The leading logarithmic impact factor $\chi^{LLA}_i$ was calculated by two of the authors directly from the Feynman diagrams in ref.~\cite{BLS2}
\beqn\label{chiLLA}
 \chi^{LLA}_1= \frac{1}{2}\frac{1}{\left(i\nu +\frac{n}{2}\right)}\left(-\frac{q_1}{k_1}\right)^{-i\nu -\frac{n}{2}}
\left(-\frac{q^*_1}{k^*_1}\right)^{-i\nu +\frac{n}{2}},\chi^{LLA}_2= -\frac{1}{2}\frac{1}{\left(i\nu -\frac{n}{2}\right)}\left(\frac{q^*_3}{k^*_2}\right)^{i\nu -\frac{n}{2}}
\left(\frac{q_3}{k_2}\right)^{i\nu +\frac{n}{2}}, \;\;\;\;\;
\eeqn
while the  NLO impact factor was extracted from (\ref{R2NLLABFKL}) and read~\cite{LP2} 
\beqn \label{NLOchi1}
\chi^{NLO}_1= \frac{a}{2}
\left(E^2_{\nu,n}-\frac{1}{4}\frac{n^2}{ \left(\nu^2+\frac{n^2}{4}\right)^2} \right)
\chi^{LLA}_1,
\eeqn
where $E_{\nu,n}$ is defined in (\ref{Enun}). The NLO correction to $\chi_2$ has a similar 
form found in ref.~\cite{LP2}. An important feature of $\chi^{NLO}_i$ in   
(\ref{NLOchi1}) is the fact that, in contrast to the leading order, it has lost the property of 
holomorphic separability: we cannot write the factor in front of $\chi^{LLA}_1$ in (\ref{NLOchi1}) as a sum of terms, which depends  only on either  $i\nu +n/2$ 
or  $-i\nu +n/2$. It is worth emphasizing 
that the NLO impact factors $\chi^{NLO}_i$ are factorized in the product of the Born 
impact factors in (\ref{chiLLA}) and a term expressed through the eigenvalue $E_{\nu,n}$ 
of the BFKL equation in the LLA. The form of the NLO impact factor in the $\nu,n$ representation 
resembles the three-loop remainder function in the LLA, emphasizing the intimate relation between the two. Indeed, it is easy to see that  expanding the integrand of (\ref{LLA}) to the third order in $a$ one gets the $E^2_{\nu,n}$ term.

In the general case the integral in (\ref{LLA}) is not easy to calculate, but 
one can consider a more restrictive kinematics, where it can be found explicitly 
at any order of the coupling $a$. One of such possibilities is the so-called collinear 
kinematics, when two adjacent particles become collinear, e.g. if in  Fig.~\ref{fig:2to4ope}  the momenta 
$p_{B}$ and $p_{B'}$ coincide. In the limit $t_3 \to 0$ the remainder function vanishes at two 
loops and beyond, in both the direct channel of Fig.~\ref{fig:2to4ope} and in the Mandelstam 
channel of Fig.~\ref{fig:2to4opeINV}. 
The multi-Regge limit followed by the collinear limit  in terms of the dual conformal cross 
ratios~(compare to the Regge kinematics in (\ref{multicross})) reads
\beqn\label{collmulticross}
1-u_1 \to +0,\;\; u_2 \to +0, \;\; u_3 \to +0, \;\; \frac{u_2}{1-u_1}=\tilde{u}_2 \to +0, \;\; \frac{u_3}{1-u_1}=\tilde{u}_3 \simeq 1,
\eeqn
which in terms of $w$ and $w^*$ implies~(see (\ref{wdef}))
 \beqn\label{collw}
1-u_1 \to +0,\;\; |w|\to +0, \;\; \cos (\phi_2-\phi_3) \simeq \mathcal{O}(1).
\eeqn
For $|w| \to 0$ the main contribution to the LLA remainder function given by (\ref{LLA}) comes from poles at $\nu =\pm i n/2$ for the conformal spin $n=1$.
In this case one can drop the $\psi$ functions in (\ref{Enun}) and perform  integration of (\ref{LLA}) at any order of the coupling constant
in the Double Leading Logarithmic Approximation~(DLLA), where one sums contributions from the powers of $a \ln |w| \ln (1-u_1)$  resulting in~\cite{BLP2}
\beqn\label{DLLAapp}
 R^{DLLA}_{2 \to 4 }=  1+
 i2\pi a \cos (\phi_2-\phi_3) \;|w| \left(1-I_0\left(2 \sqrt{a \ln |w| \ln(1-u_1)}\right) \right),
\eeqn
where $I_n(z)$ is the modified Bessel function.
The Double Leading Logarithmic Approximation is analogous to the summation of the 
contributions  of powers of $g^2 \ln s \ln Q^2$ in Deep Inelastic Scattering~(DIS), 
where the BFKL and the Dokshitzer-Gribov-Lipatov-Altarelli-Parisi~(DGLAP) equations overlap. 
Then the real part of the next-to-leading corrections  to the remainder function in DLLA was 
calculated~\cite{LP2} using the dispersion relation (\ref{disp}) and (\ref{DLLAapp}).

\subsection{$3 \to 3 $ scattering amplitude}
\begin{figure}[htbp]
	\begin{center}
		\epsfig{figure=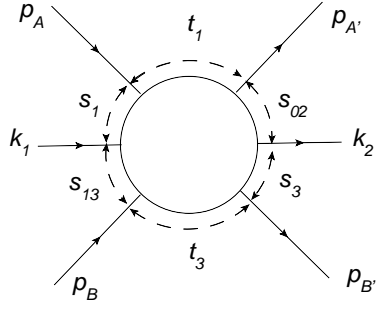,width=50mm}
	\end{center}
	\caption{ The $3 \to 3$ gluon scattering amplitude. }
	\label{fig:3to3direct}
\end{figure}

\begin{figure}[htbp]
	\begin{center}
		\epsfig{figure=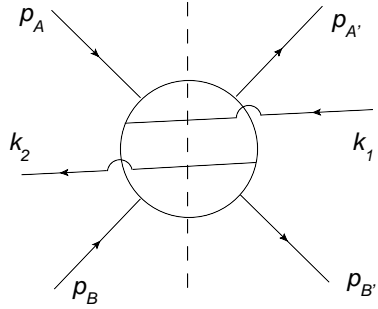,width=50mm}
	\end{center}
	\caption{ The $3 \to 3$ gluon scattering amplitude in the Mandelstam  channel given by $s_1, s_3, s_{13}, s_{02}<0\;\;\; \text{and} \;\;\; s, t^{'}_2>0$.  }
	\label{fig:3to3NOTdirect}
\end{figure}

Beyond the $2 \to 4$ scattering the six-particle scattering amplitude describes also the $3 \to 3 $ scattering illustrated in Fig.~\ref{fig:3to3direct}. The analysis of the  $3 \to 3 $ amplitude is similar to that of the $2 \to 4$, but there are some rather interesting features we want to emphasize.
Firstly we start with the definition of the  kinematic invariants  $s_{13}=(p_B+k_1)^2,  s_{02}=(p_{A'}+k_2)^2,
s=(p_B+k_1+p_A)^2, t^{'}_{2}=(p_A-p_{A'}-k_2)^2, s_1=(k_1+p_A)^2,  s_3=(p_{B'}+k_2)^2, t_2=(p_{A}-p_{A'}+k_1)^2, t_1=(p_A-p_{A'})^2$ and $t_3=(p_{B}-p_{B'})^2$. The dual conformal cross ratios are expressed in terms of these invariants as follows
\beqn\label{crossinv3to3}
u_1=\frac{s_{13}s_{02}}{s\; t^{'}_2}, \; u_2=\frac{t_{1}t_{3}}{t_2  t^{'}_2},\; u_3=\frac{s_{1}s_{3}}{s\; t_2}.
\eeqn
In the multi-Regge  kinematics for  the direct channel in Fig.~\ref{fig:3to3direct}, where all invariants are negative
\beqn
-s \gg -s_1, -s_3, -t^{'}_2 \gg -t_1, -t_2, -t_3 >0,
\eeqn
 the remainder function $R_{3 \to 3}^{(l)}$ is zero, while in the physical region of the Mandelstam  channel depicted in Fig.~\ref{fig:3to3NOTdirect}, where
\beqn\label{mandchannel}
s_1, s_3, s_{13}, s_{02} <0\;\;\; \text{and} \;\;\; s, t^{'}_2>0,
\eeqn
it contains a non-vanishing contribution, growing with energy $t^{'}_2$.
In the Mandelstam  channel~(\ref{mandchannel}) in the multi-Regge kinematics the dual conformal cross ratios (\ref{crossinv3to3}) possess a non-zero phase
\beqn\label{3to3cont}
u_1 \to |u_1| e^{i 2\pi}, \; u_2 \to |u_2| e^{i \pi},\; u_3 \to |u_3| e^{i \pi}
\eeqn
and the analytic continuation of the GSVV expression in multi-Regge kinematics gives
\beqn\label{R62cont3to3}
&& R_{3 \to 3}^{(2)\; LLA + NLLA}
 = -\frac{i\pi}{2}\ln(u_1-1)\ln|1+w|^2 \ln \left|1+\frac{1}{w}\right|^2    +\frac{\pi^2}{2}\ln|1+w|^2 \ln \left|1+\frac{1}{w}\right|^2
\\
&&
-\frac{i\pi}{2}\ln|w|^2 \ln^2|1+w|^2+\frac{i\pi}{3}\ln^3|1+w|^2 -i\pi \ln |w|^2 \left(\text{Li}_2(-w)+\text{Li}_2(-w^*)\right) \nonumber\\
&&
+i2\pi \left(\text{Li}_3(-w)+\text{Li}_3(-w^*)\right). \nonumber
\eeqn
As in the $2\to 4$ case the remainder function has the target-projectile symmetry~($p_A\leftrightarrow~p_{B}$, $p_{A'}\leftrightarrow~p_{B'}$, $k_1\leftrightarrow~k_{2}$ or $|w| \to 1/|w|$), but in contrast to the $2 \to 4$ amplitude (\ref{R62cont3to3}) has a real part $\frac{\pi^2}{2}\ln|1+w|^2 \ln \left|1+\frac{1}{w}\right|^2$.  This fact is in full agreement with the dispersion-like relation~\cite{LipDisp} for the $3\to 3$ amplitude 
\beqn\label{Reqn3to3}
R_{3 \to 3} e^{-i\pi \delta} =\cos \pi \omega_{ab} -i  \int_{-i\infty}^{i\infty} \frac{d \omega }{2\pi i}f(\omega) |1-u_1|^{-\omega}, \;\;\;\;
\eeqn
which differs from (\ref{disp}) by signs of the phase on LHS and the  integral on RHS, as well the absence of the phase $e^{-i\pi \omega}$ in the integrand.  This phase mixes the contribution from the Regge poles and the Mandelstam cut in the dispersion relation (\ref{disp}), which leads to the full cancellation of the real part at two loops for the $2 \to 4$ amplitude. In the $3 \to 3$ case this cancellation does not happen anymore due the absence of the phase $e^{-i\pi \omega}$, and one obtains a real term in (\ref{R62cont3to3}). The real part does not cancel out in both the $2 \to 4$ and $3 \to 3$ amplitudes at higher loops. The dispersion relation~(\ref{Reqn3to3}) was used by two of the authors~\cite{LP2} to find the LLA  and the real part of the NLLA contributions to the remainder function of the $3 \to 3$ amplitude
\beqn\label{R63LLA3to3}
&& \hspace{-0.5cm}R_{3 \to 3}^{(3)\; LLA}=-\frac{i\pi}{4}\ln^2(u_1-1)\left(
\ln|w|^2 \ln^2|1+w|^2
-\frac{2}{3}\ln^3|1+w|^2
\right. \hspace{0.5cm}\\
&& \left.
+\frac{1}{2} \ln|w|^2 \left(\text{Li}_2 (-w)+\text{Li}_2 (-w^*)\right)
-\frac{1}{4}\ln^2|w|^2 \ln|1+w|^2-\text{Li}_3 (-w)-\text{Li}_3 (-w^*)
\right)\nonumber
\eeqn
and
\beqn\label{R63reNLLA3to3}
&& \hspace{-0.5cm}\Re \left(R_{3 \to 3}^{(3)\; NLLA} \right)=-\frac{\pi^2}{4}\ln(u_1-1)\left( \ln^2 |1+w|^2 \ln \left|1+\frac{1}{w} \right|^2
 +\ln |1+w|^2 \ln^2 \left|1+\frac{1}{w} \right|^2 \right). \;\;\;\;
\eeqn
As we have already mentioned in the   $3 \to 3$ case there is no mixing between the real contributions coming from the Regge poles and Mandelstam cuts. This fact allows us to make a prediction valid at an arbitrary value of the coupling constant for the following expression~\cite{BLP1}
\beqn\label{allrealsub}
\Re\left(R_{3 \to 3}e^{-i\pi \delta}\right)= \cos \pi \omega_{ab}.
\eeqn
Both $\delta$ and $\omega_{ab}$ are known and are given by (\ref{deltaomega}) as  functions of the cusp anomalous dimension and the anharmonic ratios.
However there is a difficulty in understanding (\ref{allrealsub}) at the strong coupling because of the rapid oscillations as $a \to \infty$.

Similarly to the $2\to4$ case one can also consider the collinear limit $t_3 \to 0$ ($|w| \to 0$)  
preceded by the Regge limit. In this kinematics it is possible to  calculate explicitly the LLA and the real part of the NLLA $3 \to 3$ remainder function at an arbitrary number of loops~\cite{BLP2}.

In the next section we return to the discussion of section~\ref{JB} and consider a composite state of an arbitrary number of reggeized gluons. These BKP states appear in the scattering amplitudes with 8 or more  external legs.

\section{Integrability of the $n$ gluon Hamiltonian}\label{LL}

Here we discuss composite states of $n$ reggeized gluons in the
adjoint representation at large $N_c$ (cf. a similar approach for the simple case $n=2$
in ref.~\cite{BLS2}).
One can write the homogeneous BKP equation for
its wave function described by an
amplitude with amputated propagators in the form~\cite{Openchain}
\begin{equation}
H\Psi =E\Psi \,,\,\,\Delta _n=-\frac{g^2N_c}{16\pi ^2}\,E\,.
\end{equation}
Here $H$ is a redefined hamiltonian obtained after a
subtraction of the gluon Regge trajectory $\omega (t)$
containing infrared divergencies.
Namely, the Regge trajectory of the composite state
is~\cite{Openchain}
\begin{equation}
\omega _n(t)=a\left(\frac{1}{\epsilon }-\ln \frac{-t}{\mu ^2}\right)+\Delta _n\,,
\,\,a=\frac{g^2N_c}{8\pi ^2}\,,
\end{equation}
where $\Delta _n$ is the infrared stable quantity expressed in terms of the energy $E$.

The hamiltonian $H$ in the multi-color limit can be written in the
holomorphically separable form
(see \cite{Openchain})
\begin{equation}
H=h+h^{\ast }\,,\,\,h=\ln \frac{p_{1}\,p_{n}}{q^{2}}
+
\sum_{r=1}^{n-1}h^t_{r,r+1}\,,\,\,q=\sum _1^np_r\,,
\label{h8hol}
\end{equation}
where the pair hamiltonian $h^t_{r,r+1}$ is transposed to the
corresponding unamputated operator (\ref{pairham})
\begin{equation}
h^t_{r,r+1}=\ln
(p_{r}p_{r+1})+p_{r}\ln (\rho _{r,r+1})\,\frac{1}{p_r}+
p_{r+1}\ln (\rho _{r,r+1})\,\frac{1}{p_{r+1}}+2\gamma \,.
\end{equation}
It is seen from  (\ref{h8hol}) that the holomorphic hamiltonian
for the composite state in the adjoint representation differs from
the corresponding expression for the singlet case $h^{(0)}$ (\ref{pairham}) after
its transposition only by the substitution
\beq
h_{n,1}\rightarrow \ln \frac{p_1\,p_n}{q^2}\,,
\eeq
which is related to the fact, that the planar Feynman diagrams have
the topology of a strip and the infrared divergencies in the Regge
trajectories of the particles $1$ and $n$ are not compensated by the
contribution from the pair potential
energy $V_{n,1}$.

It turns out, that the eigenvalues $E$ do not depend on $|q|^2$
due to the scale invariance of $H$,
as it will be demonstrated below. As a result, the $t$-dependence of
$\omega _n(t)$ is the same as in the gluon Regge trajectory.

The normalization condition for the wave function in two-dimensional
space can be written as follows
\begin{equation}
||\Psi ||^2=\int \prod _{r=1}^{n-1}d^2p_r \,
\Psi ^*\prod _{s=1}^n|p_s|^{-2}\Psi \,,\,\,\sum _{s=1}^np_s=q\,.
\end{equation}

Using the duality transformation (see \cite{dual} and \cite{Openchain})
\begin{equation}
p_1=z_{0,1}\,,\,\,p_r=z_{r-1,r}\,,\,\,q=z_{0,n}\,,\,\,\rho _{r,r+1}=
i\frac{\partial}{\partial z_r}=i\partial _r\,,
\label{dualtr}
\end{equation}
the holomorphic hamiltonian can be presented as follows
\begin{equation}
h=\ln \frac{z_{0,1}\,z_{n-1,n}}{z_{0,n}^{2}}
+\sum_{r=1}^{n-1}h^t_{r,r+1}\,.
\label{hol1}
\end{equation}

Further, by regrouping its terms we can write the holomorphic
hamiltonian in another form~\cite{Openchain}
\begin{equation}
h=-2\ln z_{0,n}
+\ln (z_{0,1}^2\partial _1)+
\ln (z_{n-1,n}^2\partial _{n-1})+2\gamma +
\sum_{r=1}^{n-2}\,h'_{r,r+1}\,,
\label{hamMob}
\end{equation}
where the new pair hamiltonian is
\[
h'_{r,r+1}=\ln (z_{r,r+1}^2\partial _r)+
\ln (z_{r,r+1}^2\partial _{r+1})-2\ln z_{r,r+1}+
2\gamma
\]
\begin{equation}
=\ln (\partial _r)+\ln (\partial _{r+1})
+\frac{1}{\partial _r}\,\ln z_{r,r+1}\,\partial _r+
\frac{1}{\partial _{r+1}}\,\ln z_{r,r+1}\,\partial _{r+1}
+2\gamma \,.
\label{h2prime}
\end{equation}
The operator $h'_{r,r+1}$ coincides in fact after the
substitution $z_r\rightarrow \rho _r$ with
the corresponding hamiltonian in the coordinate representation
(\ref{pairham}) acting on the wave function with
non-amputated
propagators.

It is important, that $h$ (\ref{hamMob}) is invariant under the M\"{o}bius
transformations
\begin{equation}
z_k\rightarrow \frac{az_k+b}{cz_k+d}
\end{equation}
and does not contain the derivatives $\partial _0$ and $\partial _n$.
Therefore we can put
\begin{equation}
z_0=0\,,\,\,z_n=\infty \,,
\label{fixing}
\end{equation}
which leads to the simplified expression for $h$
\begin{equation}
h\rightarrow h'=
\ln (z_{1}^2\partial _1)+
\ln (\partial _{n-1} )+2\gamma +
\sum_{r=1}^{n-2}\,h'_{r,r+1}\,.
\label{hol2}
\end{equation}
To return to initial variables in the final expression for the wave function
one should perform
the following substitution of $z_k$
\begin{equation}
z_k \rightarrow -\frac{z_k-z_0}{z_k-z_n}=
\frac{\sum _{r=1}^kp_r}{q-\sum _{r=1}^kp_r}\,.
\label{substzp}
\end{equation}

According to the above representation (\ref{hol1}) for $h$, its
transposed part ${h'}^{t}$
can be obtained from $h$ by the similarity transformation which can
be written in terms of $h'$ as follows
\begin{equation}
h^{\prime \,t}=z_1^{-1}\,\left(\prod _{r=1}^{n-2}z_{r,r+1}\right)^{-1}\,
h'\,\,
z_1\,\left(\prod _{r=1}^{n-2}z_{r,r+1}\right)\,,
\end{equation}
which is compatible with the following normalization condition for the
wave function in the full two-dimensional space
\begin{equation}
||\Psi ||^2_1=\int \frac{d^2z_{n-1}}{|z_1|^2}\,\prod _{r=1}^{n-2}
\frac{d^2z _{r}}{|z_{r,r+1}|^2}\,|\Psi |^2\,.
\end{equation}

On the other hand, from the expression (\ref{hol2}) for $h'$
we obtain another relation for $h^{\prime \,t}$
\begin{equation}
h^{\prime \,t}=\left(\prod _{r=1}^{n-1}\partial _r\right)\,
h'\,\left(\prod _{r=1}^{n-1}\partial _r\right)^{-1}\,,
\end{equation}
corresponding to the second normalization condition for $\Psi$
compatible with the hermicity properties of the total hamiltonian
\begin{equation}
||\Psi ||^2_2=\int \prod _{r=1}^{n-1}
d^2z _{r}\,\Psi ^*\prod _{r=1}^{n-1}|\partial _r|^2\,\Psi \,.
\end{equation}

By comparing two above relations between $h'$ and
 $h^{\prime \,t}$ one can conclude (cf. \cite{int}), that the operator
\begin{equation}
A'=z_1\,\prod _{s=1}^{n-2}z_{s,s+1}\,
\prod _{r=1}^{n-1}\partial _r
\label{Aprime}
\end{equation}
commutes with the holomorphic hamiltonian
\begin{equation}
[A',h']=0\,.
\end{equation}

\subsection{Integrable open spin chain}

Let us verify~\cite{Openchain}, that the holomorphic hamiltonian $h'$ (\ref{hol2}) also
commutes with the
differential operator $D(u)$ being the matrix element $T_{22}$ of
the monodromy
matrix (cf.~\cite{int})
\begin{equation}
T(u)=\left(%
\begin{array}{cc}
  A(u) & B(u) \\
  C(u) & D(u) \\
\end{array}%
\right)=L_1(u)L_2(u)...L_{n-1}(u)\,,
\end{equation}
where the $L$-operator is defined by the relation
\begin{equation}
L_r(u)=\left(
\begin{array}{cc}
  u +iz_r\partial _r& i\partial _r \\
  -iz_r^2\partial _r & u-iz_r\partial  _r \\
\end{array}
\right)\,.
\end{equation}

To prove the commutativity of $h'$ and $D(u)$ one can use the
following relation~\cite{Openchain}
\begin{equation}
[L_{k}(u)\,L_{k+1}(u), h'_{k,k+1}]=
-i\left(L_{k}(u)-L_{k+1}(u)\right)\,,
\label{LLh}
\end{equation}
valid in particular  due to the M\"{o}bius symmetry of the pair hamiltonian
\begin{equation}
[\vec{M}_{k,k+1}, h'_{k,k+1}]=0\,,\,\,\vec{M}_{k,k+1}=
\vec{M}_{k}+\vec{M}_{k+1}.
\end{equation}
and the fact that its eigenvalue is a linear combination of polygamma functions~(see~(\ref{PomE})) 

Relation (\ref{LLh}) leads to the equality
\begin{equation}
[T(u),\sum_{r=1}^{n-2}h'_{r,r+1}]=
iL_2(u)L_3(u)...L_{n-1}(u)-iL_1(u)L_2(u)...L_{n-2}(u)\,.
\end{equation}
On the other hand, one can easily verify, that
\[
[T_{22}(u),\ln (z_1^2\partial _1)+\ln \partial _{n-1}]=
\left(
0 \,,\,\,1
\right)[T(u),\ln (z_1^2\partial _1)+\ln \partial _{n-1}]
\left(
\begin{array}{cc}
  0 \\
  1 \\
\end{array}
\right)
\]
\begin{equation}
=
-i
\left(
0 \,,\,\, 1
\right)\left(L_2(u)L_3(u)...L_{n-1}(u)-L_1(u)L_2(u)...L_{n-2}(u)
\right)
\left(
\begin{array}{cc}
  0 \\
  1 \\
\end{array}
\right)
\,,
\end{equation}
which proves that the differential operator $D(u)=T_{22}(u)$ is an
integral of motion~\cite{Openchain}
\begin{equation}
[D(u),h']=0\,.
\label{commrel}
\end{equation}
Note, that, if instead of condition (\ref{fixing}) we shall use the
equivalent condition
\[
z_0=\infty\,,\,\,z_n=0\,,
\]
the matrix element $A(u)=T_{11}(u)$ of the monodromy matrix will be an integral of motion.

Thus, our hamiltonian is the local hamiltonian for an open integrable
Heisenberg spin model with the spins which are generators of the
M\"{o}bius group.

With the use of the following decomposition of the $L$-operators
\begin{equation}
L_r(u)=\left(
\begin{array}{cc}
  u & 0 \\
  0 & u\\
\end{array}
\right)+
\left(
\begin{array}{cc}
  1\\
  -z_r\\
\end{array}
\right)
\left(
 z_r \,,\,\,1
\right)\,i\partial _r
\end{equation}
one can construct the matrix element $T_{22}=D(u)$ in an explicit way
\begin{equation}
D(u)=\sum _{k=0}^{n-1}u^{n-1-k}\,q'_k\,,
\end{equation}
where
\begin{equation}
q'_k=-\sum _{0<r_1<r_2<...<r_{k}<n}z_{r_1}\,
\prod _{s=1}^{k-1}z _{r_s,r_{s+1}}\,
\prod _{t=1}^k i\partial _{r_t}\,.
\end{equation}

Note, that one can parameterize the monodromy matrix in another form
\begin{equation}
T(u)=\left(
\begin{array}{cc}
  j_0(u) +j_3(u) & j_+(u) \\
  j_-(u) & j_0(u) -j_3(u) \\
\end{array}
\right)\,,\,j_\pm (u)=j_1(u)\pm ij_2(u)\,.
\end{equation}
In this case the Yang-Baxter equations for the currents $j_{\mu }$ have
the Lorentz-invariant representation~\cite{dual}
\begin{equation}
[j_\mu (u),j_\nu (v)]=\frac{\epsilon _{\mu \nu \rho \sigma}}{2(u-v)}
\left(j^\rho (u) j^\sigma (v)-j^\rho (v) j^\sigma (u)\right)\,.
\end{equation}
Here $\epsilon _{\mu \nu \rho \sigma}$ is the antisymmetric tensor
in the four-dimensional Minkowski space and
$\epsilon _{1 2 3 0}=1\,,\,\,g _{\mu \nu} =(1,-1,-1,-1)$.

\subsection{Composite states of two and three gluons}

In the case $n=2$ we have only one non-trivial
integral of motion
\begin{equation}
q'_1=-iz_1\,\partial _1 \,.
\end{equation}
Taking into account the normalization
condition for the eigenfunction in the two-dimensional space
\begin{equation}
||\Psi ||^2 =\int \frac{d^2z_1}{|z_1|^2}\,|\Psi |^2\,,
\end{equation}
we find the orthonormalized and complete set of eigenfunctions
\begin{equation}
\Psi _{m,\widetilde{m}}^{(2)}=z_1^{-\frac{1}{2}+m}\,
(z^*_1)^{-\frac{1}{2}+\widetilde{m}}\,,\,\,m=\frac{1+n}{2}+i\nu \,,\,\,
\widetilde{m}=\frac{1+n}{2}-i\nu \,,
\label{wave2}
\end{equation}
satisfying the single-valuedness requirement. Note, that
using the substitution (\ref{substzp}) one can reproduce
the wave functions of two gluon composite states  in the momentum space
(see \cite{BLS2}).

For the case $n=3$ the operator $D(u)$ is given below
\begin{equation}
D_3(u)=u^2-iu(z_1\,\partial _1+z_2 \,\partial _2)+z_1z_{1,2}\,
\partial _1\partial _2 \,.
\end{equation}
With taking into account the normalization condition
\begin{equation}
||\Psi ||^2=\int \frac{d^2z_1\,d^2z_2}{|z_1|^2|z_{1,2}|^2}\,|\Psi |^2\,,
\end{equation}
one can search
the holomorphic eigenfunction of this operator in the form
\begin{equation}
\Psi _m^{(3)}=z_2^{-\frac{1}{2}+m}\,f\left(\frac{z_2}{z_1}\right)\,.
\label{sol3m}
\end{equation}
The function $f(x)$ satisfies the equation
\begin{equation}
\left(x(1-x)\partial ^2+(\frac{1}{2}+m)(1-x)\partial +\lambda \right)
\,f=0\,,\,\,x=\frac{z_2}{z_1}\,,
\end{equation}
where $\lambda $ is the eigenvalue of the operator
$z_1z_{1,2}\partial _1\partial _2$.
Two independent solutions of this equation can be expressed in terms
of the hypergeometric function $F$
\begin{equation}\label{hyperLL}
f_1(x)=F(a_1,a_2;1+a_1+a_2;x)\,,\,\,
f_2(x)=x^{a_1+a_2}\,F(-a_2,-a_1;
1-a_1-a_2;x)\,,
\end{equation}
where the parameters $a_1$ and $a_2$ are obtained from the set of
equations
\begin{equation}
a_1+a_2=-\frac{1}{2}+m\,,\,\,a_1a_2=-\lambda \,.
\label{a1a2}
\end{equation}
We have the similar solutions for the eigenvalue of the operator $D^*$
in the antiholomorphic subspace. They can be obtained by
the substitution
\begin{equation}
x\rightarrow x^*\,,\,\,a_1\rightarrow \widetilde{a}_1\,,\,\,a_2\rightarrow \widetilde{a}_2\,,\,\,
m\rightarrow \widetilde{m}=\frac{1-n}{2}+i\nu \,.
\end{equation}
To construct the wave function $\Psi$
with the property of the single-valuedness in the two-dimensional
subspaces $\vec{z}_1$ and $\vec{x}$
we should write a bilinear combination of the functions $f_i(x)$ and the
corresponding functions in the anti-holomorphic subspace $\widetilde{f}_i(x^*)$.

One can write the integral representation for the wave function
satisfying the above constraints
\begin{equation}
\Psi \sim z^{a_1+a_2}_2\,(z^*_2)^{\widetilde{a_1}+\widetilde{a_2}}\,
\int \frac{d^2y}{|y|^2}\,
y^{-a_2}(y^*)^{-\widetilde{a_2}}\,\left(\frac{y-1}{y-x}\right)^{a_1}\,
\left(\frac{y^*-1}{y^*-x^*}\right)^{\widetilde{a_1}}\,,\,\,
x=\frac{z_2}{z_1}\,,
\end{equation}
where the integration is performed over the two-dimensional plane $\vec{y}$. Note, that
the integrand has no ambiguity in the points $y=0,1,x$ due to the additional constraints for the parameters
$a_i,\widetilde{a_i}$~\cite{Openchain}
\begin{equation}
a_1-\widetilde{a_1}=N_{a_1}\,,\,\,a_2-\widetilde{a_2}=N_{a_2}\,,
\end{equation}
where $N_{a_1},\,N_{a_2}$ are integers.
Moreover, the function $\Psi$
near the points $x=0,1, \infty$ can be presented
in terms of the sum of products of hypergeometric functions in (\ref{hyperLL}).

\subsection{Hamiltonian and integrals of motion}

The holomorphic hamiltonian for composite states of two reggeized gluons
can be written as follows
\begin{equation}
\widetilde{h}=\ln (z_1^2\partial _1)+\ln (\partial _1)+2\gamma
=\psi (z_1\partial _1)+\psi (-z_1 \partial _1)+2\gamma \,.
\end{equation}
Acting by $\widetilde{h}$ on the function $z_1^{\delta}$ we obtain
\begin{equation}
\widetilde{h}z_1^{\delta }=\epsilon (\delta )\,z_1^{\delta}\,,\,\,
\epsilon (\delta )=\psi (\delta )+\psi (-\delta )+2\gamma \,.
\end{equation}
In the case of wave function (\ref{wave2}) satisfying the
single-valuedness and
orthonormality conditions
in the two-dimensional space one derives the following expression for
the total energy~\cite{BLS2}
\begin{equation}
E_{m,\widetilde{m}}=\epsilon _m +\epsilon _{\widetilde{m}}\,,\,\,
\epsilon _m=\psi (-\frac{1}{2}+m)+
\psi (\frac{1}{2}-m)+2\gamma \,.
\end{equation}
Note, that it does not coincide with the corresponding
result (\ref{PomE}) for the Pomeron state.

For the case of  composite states of $n$ reggeized gluons the holomorphic
hamiltonian (\ref{hamMob}) in the region
\begin{equation}
z_1 \ll z_2 \ll z_3 \ll ...\ll z_{n-1}\,.
\end{equation}
becomes the sum of the disconnected hamiltonians
\begin{equation}
h'=\sum_{r=1}^{n-1}\left(\psi (z_r\partial _1)+\psi (-z_r \partial _1)+
2\gamma \right) \,.
\end{equation}
As a result, we obtain, that the wave function in this limit
is factorized~\cite{Openchain}
\begin{equation}
\Psi _{a_1,a_2,...,a_{n-1}}=\prod _{r=1}^{n-1}z_r^{a_r}
\end{equation}
and the energy is the sum of the particle energies
\begin{equation}
\epsilon =\sum _{r=1}^{n-1}\epsilon (a_r)\,.
\end{equation}
The parameters $a_r$ for these solutions and  $\widetilde{a_r}$
for anti-holomorphic solutions are obtained from the single-valuedness condition
and the normalizability
\[
a_r=i\nu _r+\frac{n_r}{2}\,,\,\,\widetilde{a_r}=i\nu _r-\frac{n_r}{2}
\,,
\]
where $\nu _r$ are real and $n_r$ are integer numbers.

The eigenvalue of the integral of motion $D(u)$ is expressed also in terms of
these parameters
\begin{equation}
\Lambda (u)=\prod _{r=1}^{n-1}(u-ia_r)\,.
\end{equation}

\subsection{The Baxter-Sklyanin approach}

To find a solution of the Yang-Baxter equation for the open spin chain one
can use the Bethe ansatz. For this purpose it is convenient to work
in the transposed representation for the monodromy matrix
\begin{equation}
T^t(u)=\left(%
\begin{array}{cc}
  j^t_0(u)+j^t_3(u) & j^t_+(u) \\
  j^t_-(u) & j^t_0(u)-j^t_3(u) \\
\end{array}%
\right)=L^t_{1}(u)L^t_{2}(u)...L^t_{n-1}(u)\,,
\end{equation}
where the $L$-operator can be chosen as follows
\begin{equation}
L^t_r(u)=\left(
\begin{array}{cc}
  u +i\partial _rz_r& i\partial _r \\
  -i\partial _rz_r^2 & u-i\partial  _r z_r\\
\end{array}
\right)\,.
\end{equation}

The pseudo-vacuum state is defined as a solution of the equation
\begin{equation}
j^t_-(u)\Psi _0=0\,.
\end{equation}
It can be written in the form~\cite{FK}
\begin{equation}
\Psi _0=\prod_{r=1}^{n-1}z_r^{-2}\,.
\end{equation}
Note, that the function $|\Psi _0|^2$ does not belong to the principal
series of the unitary representations. As a result, the states constructed
in the framework of the Bethe ansatz by applying  the product
of the operators $j^r_+(u_r)$ to $\Psi _0$
\begin{equation}
\Psi ^t_k =\prod _{r=1}^kj^t_+(u_r)\,\Psi _0
\end{equation}
are non-physical. Nevertheless, these states are eigenfunctions of
the integral of motion
\begin{equation}
D^t(u)\Psi ^t _k=(j^t_0(u)-j^t_3(u))\Psi ^t_k=\Lambda (u)\Psi ^t_k
\end{equation}
providing that
\begin{equation}
\Lambda (u)=(u+i)^{n-1}\prod _{t=1}^k\frac{u-u_t+i}{u-u_t} \equiv
(u+i)^{n-1}\frac{Q(u+i)}{Q(u)}
\end{equation}
is a polynomial, which leads to a quantization condition for
the Bethe roots $u_t$. If we parameterize this polynomial as follows
\begin{equation}
\Lambda (u)=\prod _{l=1}^{n-1}(u-ia_l)\,,
\end{equation}
the above defined Baxter function $Q(u)$ can be calculated
\begin{equation}
Q(u)=\prod _{l=1}^{n-1}\frac {\Gamma (-iu-a_l)}{\Gamma (-iu+1)}\,.
\end{equation}

As it was mentioned above, the polynomial solutions for $Q(u)$ are non-physical,
because the
corresponding wave functions $\Psi $ do not belong to the principal
series of unitary representations of the M\"{o}bius group. We should find
a set of non-polynomial solutions $Q_{s}(u)$ satisfying this physical
requirement.

According to E. Sklyanin~\cite{Sklya} the correct variables in which the
dynamics of
the Heisenberg spin model is drastically simplified are the zeroes
$\hat{b}_r$ of the
operator $B(u)=j^t_+(u)$ entering in the monodromy matrix
\begin{equation}
B(u)=P_{n-1}\,\prod _{k=1}^{n-2}(u-\hat{b}_r)\,,\,\,
P_{n-1}=i\sum _{r=1}^{n-1}\partial _r \,,
\end{equation}
where the operators $\hat{b}_r$ and $P_{n-1}$
commute each with others
\begin{equation}
[\hat{b}_r,\hat{b_s}]=[\hat{b}_r,P_{n-1}]=0\,.
\end{equation}

It is convenient to pass from the coordinate representation $\vec{z}$ to
the Baxter-Sklyanin representation~\cite{dVL}, in which the currents
$j^t_+(u)$ and $(j^t_+(u))^*$
(together with the operators
$\hat{b}_r, \hat{b}_r^*$ and $P_{n-1}, P_{n-1}^*$)
are diagonal. We denote the eigenvalues of the Sklyanin operators by
$b_r, b_r^*$.
The kernel of the unitary transformation to the Baxter-Sklyanin representation
is known explicitly for the cases $n=2$, $n=3$ and $n=4$~\cite{dVL}.
For general $n$ this integral operator can be presented as a
multi-dimensional integral~\cite{DKM}.

In the Baxter-Sklyanin representation  the wave function in the
holomorphic subspace can be expressed as a
product of the pseudo-vacuum state in this representation
$\Psi _0(P_{n-1},b_1,b_2,...,b_{n-2})$ and the Baxter
functions $Q(u_t)$
\begin{equation}
\Psi ^t(P_{n-1};b_1,...,b_{n-2})=
P _{n-1}^{-\frac{n-1}{2}-m}\prod _{k=1}^{n-2}Q(b_k)\,
\Psi _0(P_{n-1},b_1,...,b_{n-2})\,,
\end{equation}
where the power-like behavior in the variable $P_{n-1}$ is in an
agreement with the normalization condition.

The analogous representation is valid for the total wave function
\begin{equation}
\Psi ^+(\vec{P}_{n-1};\vec{b}_1,...,\vec{b}_{n-2})=
P _{n-1}^{-\frac{n-1}{2}-m}\,(P^* _{n-1})^{-\frac{n-1}{2}-\widetilde{m}}
\prod _{k=1}^{n-2}Q(\vec{b}_r)\,
\Psi _0(\vec{P}_{n-1};\vec{b}_1,...,\vec{b}_{n-2})
\end{equation}
with the use of the generalized
Baxter function $Q(\vec{u})$ being a bilinear combination of the usual
Baxter functions in the holomorphic and anti-holomorphic subspaces
\begin{equation}
Q(\vec{u})=\sum _{s,t}d_{s,t}\,Q_{s}(u)\,Q_t(u^*)\,.
\end{equation}
Here $Q_{s}(u)$ are different solutions of the Baxter equation with
the same eigenvalue $\Lambda (u)$. The coefficients $d_{s,t}$ are chosen
from the requirement, that the function $Q(\vec{u})$ satisfies the
normalization condition everywhere including the points where the
functions $Q_{s}(u)$ and $Q_t(u^*)$ have the poles~\cite{dVL, DKM}.
For the periodic spin chain this condition leads to the
quantization of the eigenvalue of the operator $A(u)+B(u)$ although
a simpler method of quantization is based on the
requirement, that all Baxter functions corresponding to the
same eigenvalue should have the same holomorphic energy~\cite{dVL}. In
the case of the open Heisenberg spin model the situation is simpler
and will be discussed below.

\subsection{Baxter-Sklyanin representation for   three gluon states}
For the states composed from three reggeized gluons the
transposed integral
of motion in the holomorphic subspace is
\begin{equation}
D^t_3(u)=u^2-iu(\partial _1\,z_1+\partial _2\,z_2)+
\partial _1\partial _2 \,z_1z_{1,2}
\end{equation}
and the operator $j_+^t$ is given below
\begin{equation}
j_+^t=iu(\partial _1+\partial _2)-\partial _1\partial _2z_{12}=
i(\partial _1+\partial _2)\,(u-\hat{b}_1)\,,
\end{equation}
where
\begin{equation}
\hat{b}_1=-i\frac{\partial _1\partial _2}{\partial _1+\partial _2}
\,z_{12}\,.
\end{equation}
The operator $j_+^t$ is easily diagonalized after a transition to
the momentum representation, where
\begin{equation}
i\partial _1\,f_{p_1,p_2}=p_1\,f_{p_1,p_2}\,,\,\,
i\partial _2\,f_{p_1,p_2}=p_2\,f_{p_1,p_2}\,.
\end{equation}
In this case the eigenvalue equation for $j_-^t$ has the form
\begin{equation}
\left(u(p_1+p_2)-i\,p_1p_2(\frac{\partial }{\partial p_1}-
\frac{\partial }{\partial p_2})\right)f =(p_1+p_2)(u-b_1)\,f
\,,
\end{equation}
where $b_1$ is the eigenvalue of $\hat{b}_1$.
Its solution  is given below
\begin{equation}
f =\chi (p_1+p_2, b_1)\,\left(\frac{p_1}{p_2}\right)^{-ib_1}\,,
\end{equation}
where $\chi $ is an arbitrary function of $p_1+p_2$ and $b_1$.
The dependence of $\Psi ^t$ from $p_1+p_2$ is fixed by the normalization
condition
\begin{equation}
\Psi ^t\sim (p_1+p_2)^{-a_1-a_2}\,.
\end{equation}

On the other hand, the eigenvalue equation for the integral of motion in
the momentum space can be written
in the form
\begin{equation}
p_1p_2\frac{\partial }{\partial p_1}\,\left(
\frac{\partial }{\partial p_2}-
\frac{\partial }{\partial p_1} \right)\Psi (p_1,p_2)=
a_1a_2 \Psi (p_1,p_2)\,.
\end{equation}
Using the ansatz
\begin{equation}
\Psi (p_1,p_2)=(p_1+p_2)^{-a_1-a_2}\,\eta
(y)\,,\,\,y=\frac{p_2}{p_1}\,,
\end{equation}
we obtain the following equation for the function $\eta(y)$
\begin{equation}
\left(y^2\,\partial ^2+(a_1+a_2+1)\,y\,\partial -a_1a_2\right)\eta (y)=
\left(-y^3\,\partial ^2-2\,y^2\,\partial \right)\eta (y)\,.
\end{equation}
There are two independent solutions of this equation~\cite{Openchain}
\begin{equation}
\eta _1 (y)=
\frac{1}{y}\,F(1-a_1,1-a_2,2;-\frac{1}{y})
\end{equation}
and
\begin{equation}
\eta _2 (y)
= -\frac{\Gamma (-a_1)\,\Gamma (+a_2)}{\Gamma (1+a_2-a_1)}
\,y^{-a_1}\,F(-a_1,1-a_1,1+a_2-a_1; -y)\,.
\end{equation}
One can construct the bilinear combination of these solutions having
the single-valuedness property in the $\vec{y}$-space.
Finally with the use of the integral representation for the hypergeometric
function the wave function $\Psi ^t $ in the momentum space can be written
as follows~\cite{Openchain}
\begin{equation}
\Psi ^t(\vec{p}_1,\vec{p}_2 )=(p_1+p_2)^{-a_1-a_2}
(p_1^*+p_2^*)^{-\widetilde{a}_1-\widetilde{a}_2}
\,\phi (\vec{y})\,,
\end{equation}
where $\phi (\vec{y})$ is given below
\begin{equation}
\phi (\vec{y})=\int d^2t\,
\left(\frac{1}{t\,y}+1\right)^{a_1}\,
\left(\frac{1}{t^*\,y^*}+1\right)^{\widetilde{a}_1}\,
(1-t)^{a_2-1}\,(1-t^*)^{\widetilde{a}_2-1}
\end{equation}
and satisfies the single valuedness condition in the $\vec{y}$-space
due to the quantization conditions for $a_r$ and $\widetilde{a_r}$.

The transition to the Baxter-Sklyanin representation
$(u,\widetilde{u})$ corresponds to the
Mellin-type transformation of $\phi (\vec{y})$
\begin{equation}
\phi (u,\widetilde{u})=\int \frac{d^2y}{|y|^2} \,y^{-iu}\,
(y^*)^{-i\widetilde{u}}
\phi (\vec{y})\,.
\end{equation}

The inverse transformation corresponds to the
Baxter-Sklyanin representation for the wave function
\begin{equation}
\Psi ^t(\vec{p}_1,\vec{p}_2 )=(p_1+p_2)^{-a_1-a_2}
(p_1^*+p_2^*)^{-\widetilde{a}_1-\widetilde{a}_2}
\,\int d^2u\,
\phi (u,\widetilde{u})\,\left(\frac{p_1}{p_2}\right)^{-iu}\,
\left(\frac{p_1^*}{p_2^*}\right)^{-i\widetilde{u}}
\,,
\end{equation}
where
\begin{equation}
-iu=i\nu _u+\frac{N_u}{2}\,,
\,\,-i\widetilde{u}=i\nu _u-\frac{N_u}{2}\,\,,\,\,\,
\int d^2u\equiv \int _{-\infty}^{\infty} d\nu _u \sum
_{N_u=-\infty}^{\infty} \,.
\end{equation}

One can interpret the wave function $\phi (u,\widetilde{u})$ in
the Baxter-Sklyanin
representation as a product of the pseudo-vacuum state
$u\,\widetilde{u}$ and the total Baxter function~\cite{Openchain}
\begin{equation}
\phi (u,\widetilde{u})=u\,\widetilde{u}\,Q(u,\widetilde{u})\,,
\end{equation}
where
\begin{equation}
Q(u,\widetilde{u})\sim
\frac{\Gamma (iu)\Gamma (i\widetilde{u})}{\Gamma (1-iu)\,
\Gamma (1-i\widetilde{u})}\,\frac{\Gamma (-iu-a_1)
\,\Gamma (-iu-a_2)}{
\Gamma (1+i\widetilde{u}+\widetilde{a}_1)
\Gamma (1+i\widetilde{u}+\widetilde{a}_2)}\,.
\end{equation}
This expression  can be written in the factorized form
\begin{equation}
Q(u,\widetilde{u})\sim Q(u,a_1,a_2)\,Q(\widetilde{u},\widetilde{a}_1,
\widetilde{a}_2)\,,
\end{equation}
where
\begin{equation}
Q(u,a_1,a_2)=
\frac{\Gamma
(-iu-a_1)\Gamma (-iu-a_2)}{\Gamma ^2(1-iu)}\,\Phi (u)\,,
\end{equation}
\begin{equation}
\Phi (u)=
\sqrt{\frac{\sin (\pi (-iu-a_1))\,\sin (\pi (-iu-a_2))}{\sin ^2(-i\pi u)}}
\,.
\end{equation}
The expression  $Q(u,a_1,a_2)$ differs from the Baxter function
in the holomorphic space
\begin{equation}
Q(u)=\frac{\Gamma
(-iu-a_1)\Gamma (-iu-a_2)
}{\Gamma
^2(1-iu)}
\end{equation}
only by the periodic function $\Phi (u)$
and therefore it can be considered also as a
Baxter function.
This additional factor 
can be included in the definition of a new pseudo-vacuum state.
Really this pseudo-vacuum state can be considered as the additional
factor for the wave function in the Baxter-Sklyanin representation
providing correct hermicity properties of
the hamiltonian and integrals of motion in
this representation\footnote{We thank Prof. F. Smirnov for discussions
related to
this important interpretation of the pseudo-vacuum state.
} (see also ref.~\cite{DKM}). 

\section{Conclusion}
In this review article we have outlined the role of Mandelstam-cut contributions 
in the remainder functions for the BDS amplitudes. Particular emphasis has been      
given to the integrability of the Hamiltonian which describes the energy spectrum 
of the states of $n$ reggeized gluons. These cut contributions appear in multi-Regge
kinematics in special physical regions, where some energies are negative.
For the cut corresponding to the composite states of $n$ reggeized gluons the number of external
particles should be  $k\ge 2+2n$. The wave functions of these
states in the adjoint representation satisfy the BFKL-like equations, which have the property of holomorphic factorization and are integrable
in LLA.
The corresponding holomorphic hamiltonian coincides with the local hamiltonian for
an integrable open Heisenberg spin model. The Baxter equation for this model
is reduced to a simple recurrence relation and can be solved in terms
of the product of the $\Gamma$-functions. We constructed the wave functions
of composite states of 2 and 3 gluons explicitly.
 \\

We thank L.~D.~Faddeev, V.~S.~Fadin, E.~M.~Levin, J.~Maldacena, A.~Sabio~Vera, V.~Schomerus, A.~Sever, M.~Spradlin, Chung-I~Tan, C.~Vergu, P.~Vieira and A.~Volovich   for helpful discussions.

\end{document}